\documentclass[%
reprint,
superscriptaddress,
amsmath,amssymb,
aps,
prapplied,
]{revtex4-2}

\usepackage[colorlinks, citecolor=blue, linkcolor=blue]{hyperref}
\usepackage{placeins} 

\usepackage{amsmath}
\usepackage{float}
\usepackage{graphicx}
\usepackage{dcolumn}
\usepackage{textcomp}
\usepackage[dvipsnames]{xcolor}
\usepackage{bm}
\usepackage{orcidlink}
\usepackage{siunitx}
\usepackage{comment}

\begin{document}

\preprint{APS/123-QED}



\title{Deterministic control of the probabilistic phase dynamics in injection-locked spin-torque nano-oscillators}

\author{Abderrazak Hakam\,\orcidlink{0009-0006-6555-8840}} 
\email{abderrazak.hakam@outlook.com}
\affiliation{Univ. Grenoble Alpes, CEA, CNRS, Grenoble INP, SPINTEC, 38000 Grenoble, France}

\author{Chloé Chopin\,\orcidlink{0000-0002-9115-3330}} 
\affiliation{Univ. Grenoble Alpes, CEA, CNRS, Grenoble INP, SPINTEC, 38000 Grenoble, France}

\author{Nhat-Tan Phan\,\orcidlink{0000-0002-2806-1660}} 
\affiliation{Univ. Grenoble Alpes, CEA, CNRS, Grenoble INP, SPINTEC, 38000 Grenoble, France}

\author{Nicolas Mollard\,\orcidlink{}}
\affiliation{Univ. Grenoble Alpes, CEA, CNRS, Grenoble INP, SPINTEC, 38000 Grenoble, France}

\author{Franck Badets\,\orcidlink{0000-0002-8331-4042}}
\affiliation{Univ. Grenoble Alpes, CEA Leti, 38000 Grenoble France}

\author{Louis Hutin\,\orcidlink{0000-0001-6429-3867}}
\affiliation{Univ. Grenoble Alpes, CEA Leti, 38000 Grenoble France}

\author{Luana Benetti\,\orcidlink{0000-0003-3063-957X}}
\affiliation{International Iberian Nanotechnology Laboratory (INL), 4715-31 Braga, Portugal}

\author{Alex S. Jenkins\,\orcidlink{0000-0002-6188-7755}}
\affiliation{International Iberian Nanotechnology Laboratory (INL), 4715-31 Braga, Portugal}

\author{Ricardo Ferreira\,\orcidlink{0000-0003-0953-2225}}
\affiliation{International Iberian Nanotechnology Laboratory (INL), 4715-31 Braga, Portugal}

\author{Ursula Ebels\,\orcidlink{0000-0001-5061-5538}}
\email{ursula.ebels@cea.fr} 
\affiliation{Univ. Grenoble Alpes, CEA, CNRS, Grenoble INP, SPINTEC, 38000 Grenoble, France}

\author{Philippe Talatchian\,\orcidlink{0000-0003-2034-6140}}
\email{philippe.talatchian@cea.fr}
\affiliation{Univ. Grenoble Alpes, CEA, CNRS, Grenoble INP, SPINTEC, 38000 Grenoble, France}

\begin{abstract}
Spin-torque nano-oscillators (STNOs) inherently exhibit thermally driven phase fluctuations that render their dynamics truly stochastic. Here, we demonstrate that, despite this intrinsic randomness, the probability of occupying each phase state can be deterministically and continuously programmed. We experimentally investigate a vortex-based STNO operating under second-harmonic injection-locking, where the oscillator phase settles into two degenerate attractors separated by $\pi$ and undergoes thermally activated phase jumps. By applying a weak radio-frequency perturbation at the free-running frequency, we tune the phase-jump rates between the two attractors without suppressing the fluctuations, achieving continuous probability control from the unbiased limit to values approaching 0 or 1. The bias phase selects which attractor is favored while the bias amplitude sets the strength of the imbalance, providing two complementary control knobs within a single nanoscale device. A phase-reduced description based on an effective quasipotential quantitatively accounts for the observations. These results establish injection-locked STNOs as programmable stochastic elements and provide a hardware primitive for probabilistic computing, Ising machines, and brain-inspired computing architectures.
\end{abstract}

\maketitle

\section{Introduction}
Spin-torque nano-oscillators (STNOs) have emerged as highly tunable, magnetic microwave components with low power consumption and pronounced nonlinearity at the nanoscale~\cite{kiselev_microwave_2003, rippard_direct-current_2004}. These oscillators exhibit wide frequency agility, making them well-suited for broadband microwave communication~\cite{choi2014spin, litvinenko2022ultrafast, zhu2023nonlinear} and unconventional computing schemes~\cite{grollier2020neuromorphic, finocchio2021promise}. Among various STNO implementations, those based on magnetic tunnel junction (MTJ) structures are particularly appealing due to their large output signal and their similarity to memory cells used in magnetic non-volatile memories, enabling complementary metal-oxide-semiconductor (CMOS) compatibility, therefore, scalable integration into microelectronic systems~\cite{ma2021microwave}.


The application of an electrical current to an STNO produces a spin-polarized electron flow across the tunnel barrier, which in turn exerts a spin-transfer torque on the free layer magnetization~\cite{berger, slonczewski, ralph2008spin}. When the applied current density exceeds a critical threshold, the spin-transfer torque drives self-sustained oscillation of the free-layer magnetization, resulting in microwave resistance oscillations through magnetoresistive effects. Importantly, the oscillation frequency of STNOs can be finely tuned by adjusting the direct current or the external magnetic field~\cite{slavin_nonlinear_2009}. In the auto-oscillation regime, STNOs can synchronize to external microwave signals through injection-locking~\cite{rippard2005injection, quinsat2011injection, dussaux2011phase, hamadeh2014perfect}, matching their oscillation frequency with that of the input within a characteristic bandwidth known as the injection-locking range~\cite{pikovsky2002synchronization}.

Injection-locking has been widely employed to suppress phase noise in synchronized STNOs~\cite{tamaru2015extremely, kreissig2017vortex, wittrock2021stabilization}. However, at room temperature, the coherence of magnetization oscillations remains susceptible to thermal fluctuations, leading to stochastic dephasing events known as phase jumps~\cite{martins2023second}. These jumps degrade the phase coherence of the oscillator and can limit performance in practical applications such as stable microwave sources. In contrast to noise-minimization efforts, our approach seeks to harness this inherent stochasticity of STNO phase dynamics. 

Recently, in Ref.~\cite{phan2024unbiased}, we introduced and demonstrated a method to generate unbiased random bitstreams by exploiting the intrinsic phase noise in STNOs, offering a compact and efficient solution. While unbiased randomness is essential for cryptographic and security-oriented applications, many unconventional computing architectures---such as Ising machines~\cite{mohseni2022ising, mcmahon2016fully, inagaki2016coherent, dutta2021ising, honjo2021100, iftakher2026intrinsic, aadit2022massively} and stochastic neural networks~\cite{mizrahi2018neural, Daniels2019EnergyefficientSC, zand2018low, kaiser2022hardware}---require fine control over the bias of random bitstreams. This constraint is typically addressed at the level of individual computing primitives known as probabilistic bits ($p$-bits)~\cite{faria2017low, sutton2017intrinsic, camsari2017stochastic, chowdhury2023full}, which have been the focus of intense hardware development using superparamagnetic tunnel junctions~\cite{pervaiz2017hardware, borders2019integer, soumah2025entropy} and their CMOS-integrated counterparts~\cite{hutin2022importance, danouchi2023designing, singh2024cmos, iftakher2026intrinsic}. In parallel, STNOs have recently attracted significant attention as building blocks for phase-based Ising machines~\cite{wang2019oim}, where the binary spin variable is encoded in the oscillator phase rather than in a magnetization state~\cite{raimondo2025high, raimondo2025adaptive}. In such architectures, the ability to deterministically tune the probability of each oscillator occupying a given phase state is essential for encoding local fields and problem constraints. Yet, to date, no experimental demonstration of an STNO operating as a $p$-bit---that is, a single device whose output probability can be continuously programmed---has been reported. 
Very recently, it has been proposed theoretically that generic oscillator Ising machines can be configured as $p$-bit engines~\cite{ekanayake2025bridging, dutta2021ising}, although no experimental demonstration with a nanoscale spintronic oscillator has been reported to date.

In this work, we experimentally demonstrate deterministic control of the stochastic phase dynamics of an injection-locked STNO, enabling precise tuning of the output bitstream probability, as verified by statistical analysis. By exploiting a weak RF bias excitation to programmably shape the phase-state occupation probabilities, we realize tunable randomness at the level of a single nanoscale oscillator. These results establish a simple and compact spintronic hardware route toward the implementation of $p$-bits and tunable stochastic binary neurons.


Section~\ref{II} introduces the basic principle of deterministic probability control in injection-locked STNOs, together with the experimental setup and the procedure used to extract dwell times and occupation probabilities. Section~\ref{III} and Section~\ref{IV} present the experimental and numerical demonstration of deterministic control of the mean dwell times and phase-state probabilities, respectively. Section~\ref{V} develops a phenomenological quasipotential model based on a modified Adler-like equation, which clarifies the physical origin of the probability and mean dwell time control and accounts for the main experimental and numerical observations. Section~\ref{VI} discusses the implications of these results for probabilistic and unconventional computing and concludes the paper. Appendices~\ref{A.A}--\ref{A.G} provide supporting details on the device, simulations, analytical derivations, data analysis, and fitting procedure.

\section{Controlling Stochastic Phase Dynamics with an RF Bias Excitation} \label{II}

\subsection{Basic principle} \label{II.A}

\begin{figure} \centering
  \includegraphics[width=85mm]{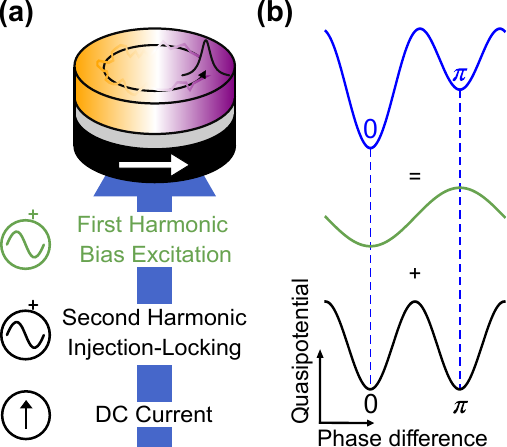}
  \caption{(a) Schematic of the experimental implementation. A vortex-based STNO is driven into auto-oscillation at frequency $f_0$ by a DC current $I_{\mathrm{DC}}$ and a static mostly out-of-plane magnetic field $H_{\mathrm{DC}}$. An RF drive at $f_{\mathrm{IL}} \approx 2f_0$ is used for second-harmonic injection-locking, and a weaker RF bias excitation at $f_{\mathrm{bias}} \approx f_0$ is applied.
  (b) Qualitative quasipotential for the phase difference $\Delta\phi$. Under the SHIL drive alone, the quasipotential is symmetric, with two equivalent minima at $0$ and $\pi$ (black). The RF bias adds a phase-dependent contribution (green), yielding a tilted quasipotential (blue) with unequal transition barriers.}
  \label{fig:1}
\end{figure}


Fig.~\ref{fig:1}~(a) illustrates the operating principle. A DC current $I_{\mathrm{DC}}$ induces auto-oscillations of the free-layer magnetization of the STNO at a free-running frequency $f_0$. A static magnetic field $H_{\mathrm{DC}}$ is applied to stabilize and adjust the oscillations, but is not essential to the underlying principle.
An RF drive at frequency $f_{\mathrm{IL}} \approx 2f_0$ is then applied, bringing the STNO into the second-harmonic injection-locking (SHIL) regime, in which the oscillator frequency is locked to half the drive frequency. Additionally, an RF bias excitation is applied at a frequency $f_{\mathrm{bias}} \approx f_0$, with an amplitude much smaller than that of the SHIL drive, so that it can be considered as a weak perturbation. 
The RF bias excitation and the SHIL drive are represented, respectively, as an in-plane RF magnetic field and an RF voltage, in accordance with the experimental implementation used in this work, but these choices are not intrinsic to the underlying principle. Likewise, the oscillator is represented as a vortex-based STNO, consistent with the experimental implementation used in this work. Such devices provide a convenient platform because of their relatively large microwave emission and well-understood dynamics. However, the same principle extends to other STNOs and nonlinear oscillators.

In the SHIL regime, the oscillator phase $\phi_{\mathrm{STNO}}$ maintains a quasi fixed relation with the SHIL drive phase $\phi_{\mathrm{IL}}$, which we choose to describe using the phase difference $\Delta\phi=\phi_{\mathrm{STNO}}-\phi_{\mathrm{IL}}/2$.
At zero temperature, $\Delta\phi$ adopts one of two stable phase configurations, $\Delta\phi\equiv0$ or $\Delta\phi\equiv\pi$ which are expressed as modulo $2\pi$, and defined up to a constant phase offset $\phi_0$. This offset is set by the locking conditions, the oscillator nonlinearity~\cite{slavin_nonlinear_2009, litvinenko2021analog, hem2019power}, and the measurement circuitry. For simplicity, we refer to these two configurations as the 0 and $\pi$ states, respectively. This situation is schematically illustrated in Fig.~\ref{fig:1}~(b) by the symmetric quasipotential induced by the SHIL drive, whose two equivalent minima correspond to the $0$ and $\pi$ states.
At finite temperature, thermal fluctuations perturb the magnetization dynamics of the free-layer and can drive the system across the barriers separating the two metastable states, thereby inducing stochastic back-and-forth phase-jumps between $0$ and $\pi$ states.
Because the corresponding transition barriers are symmetric, the phase-jump rates are equal, and the probability of occupying either state is the same (0.5), yielding an unbiased switching process, as demonstrated in our previous work in Ref.~\cite{phan2024unbiased}.

To break this symmetry, the RF bias excitation is introduced, adding a $2\pi$ contribution to the quasipotential, shown schematically by the green curve in Fig.~\ref{fig:1}~(b). When this contribution is added to the SHIL quasipotential, it produces the asymmetric quasipotential shown in blue. The barriers for the $0 \rightarrow \pi$ and $\pi \rightarrow 0$ phase-jumps then become different. As a result, the two phase-jump rates are no longer equal, and the occupation probabilities of the $0$ and $\pi$ states become unequal ($\neq 0.5$). As we show in Sec.~\ref{III}, this occupation probability can be tuned continuously through the amplitude of the bias excitation and, most importantly, through its phase $\alpha$ relative to the SHIL drive. In this way, $\Delta\phi$ continues to undergo stochastic back-and-forth switching because of thermal fluctuations, while the probability of finding the system in the $0$ or $\pi$ state becomes deterministically programmable using the RF bias excitation.

\subsection{Experimental setup} \label{II.B}

\begin{figure} \centering
  \includegraphics[width=85mm]{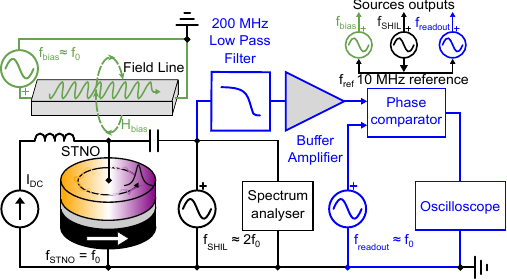}
  \caption{Experimental circuit schematic. A DC current source injects a constant current into the STNO, inducing vortex-core auto-oscillations at frequency $f_0$. Second-harmonic injection-locking is achieved by applying an RF voltage at $f_{\mathrm{IL}} \approx 2f_0$ through a bias tee. An RF bias excitation at $f_{\mathrm{bias}} \approx f_0$ is applied to a field-line fabricated above the device, generating a local in-plane RF magnetic field that influences the vortex-core dynamics. The emitted RF signal is analyzed either in the frequency domain with a spectrum analyzer or in the time domain with an analog phase-readout circuit comprising a 200~MHz low-pass filter, a 12~dB buffer amplifier, a phase comparator referenced to an additional RF source at $f_{\mathrm{readout}} = f_{\mathrm{IL}}/2$, and an oscilloscope. The RF sources used for the SHIL drive, the bias excitation, and the readout reference are all synchronized to the same 10~MHz reference, ensuring a common phase reference throughout the experiment.
  }
  \label{fig:2}
\end{figure}

To investigate the stochastic phase dynamics of STNOs in the SHIL regime, we developed an electrical setup that combines frequency-domain characterization with real-time phase readout, as shown in Fig.~\ref{fig:2}. Here, the setup is tailored to a vortex-based STNO operating in the sub-GHz range, although the same circuit can be readily extended to higher-frequency ranges. We use a DC current source to inject a current $I_{\mathrm{DC}}$ into the STNO, while a static magnetic field $H_{\mathrm{DC}}$, mostly out-of-plane,  is applied to stabilize the vortex-based auto-oscillating regime. Under these conditions, the vortex core undergoes gyrotropic motion in the plane of the free layer, giving rise to RF magnetoresistance auto-oscillations at the free-running frequency $f_0$. To drive the oscillator into the SHIL regime, we apply to the STNO an RF voltage generated by a microwave source at frequency $f_{\mathrm{IL}} \approx 2f_0$. We use a bias-tee to combine the injected DC current and RF drive at the STNO, while separating the measured RF signal from the DC component. An additional RF voltage source (green) is used to apply a bias excitation at frequency $f_{\mathrm{bias}} = f_{\mathrm{IL}}/2 \approx f_0$ through a field-line antenna fabricated above the STNO free layer. This excitation generates, at the same frequency, a local in-plane RF magnetic field at the position of the magnetic vortex. The resulting RF field amplitude is controlled by the applied ``bias power" $\mathrm{P}_\mathrm{bias}$, and its ``bias phase" $\alpha$ set by the RF voltage source. 

The collected RF signal, which includes the STNO emission, is then routed into two parallel measurement paths, one to a spectrum analyzer for frequency-domain characterization and the other to the analog phase-readout circuit shown in blue in Fig.~\ref{fig:2}~(a), enabling direct real-time access to the stochastic STNO phase dynamics in the time-domain. In this read-out circuit, the STNO signal first passes through a 200~MHz low-pass filter that eliminates SHIL drive signal ($\approx 2f_0$).  Although the free-running frequency of the STNO ($f_0\approx243.76$~MHz) lies near the filter's frequency cutoff, within the transition band, the finite roll-off of the filter ensures the STNO's emission is isolated with negligible attenuation. The filtered signal is then sent through a buffer amplifier with an effective gain of 12~dB, used primarily to prevent back reflections from the downstream electronics that could affect the STNO dynamics. In our experiments, the RF emission of the vortex-based STNO is sufficiently strong that this stage is included mainly for isolation rather than amplification.

After the buffer stage, the signal is sent to a phase comparator circuit, where it is compared with a reference signal generated by an additional RF source at frequency $f_{\mathrm{readout}} = f_{\mathrm{IL}}/2$. Using this reference signal, the phase comparator outputs a voltage proportional to the phase difference $\Delta\phi = \phi_{\mathrm{STNO}} - \phi_{\mathrm{IL}}/2$, which is subsequently recorded on a single-shot oscilloscope, thereby providing direct access to the time evolution of the stochastic phase dynamics of the STNO. 

As multiple signal phases are manipulated in the experiment, they must all be defined with respect to a common phase reference. Accordingly, the RF voltage sources used to generate the SHIL drive, the RF bias excitation, and the readout reference signal are all synchronized to the same internal 10~MHz reference. In this way, $\alpha$ can be controlled directly through the phase of the source generating the RF bias excitation, without additional phase-offset correction.

The results presented in Sec.~\ref{II}~(C-D), and Sec.~\ref{III} were obtained with the DC current set to $I_{\mathrm{DC}}=-4.4$~mA and the applied magnetic field measured to be $\mu_0H_{\mathrm{DC}}\approx 81.3$~mT, mainly out of plane with slight in-plane components, leading to a free-running frequency of $f_0\approx 243.76$~MHz. The SHIL drive was applied at $f_{\mathrm{IL}}=488$~MHz with a power $P_{\mathrm{IL}}=-9$~dBm. Table~\ref{tab:A.B.1} in Appendix~\ref{A.B} summarizes the oscillator frequency, linewidth, and integrated power extracted from Lorentzian fits in the free-running and SHIL regimes.

\subsection{Experimental demonstration of deterministic 0 or \texorpdfstring{\boldsymbol{$\pi$}}{pi}-phase-state selection} \label{II.C} 

\begin{figure}[ht] \centering
  \includegraphics[width=85mm]{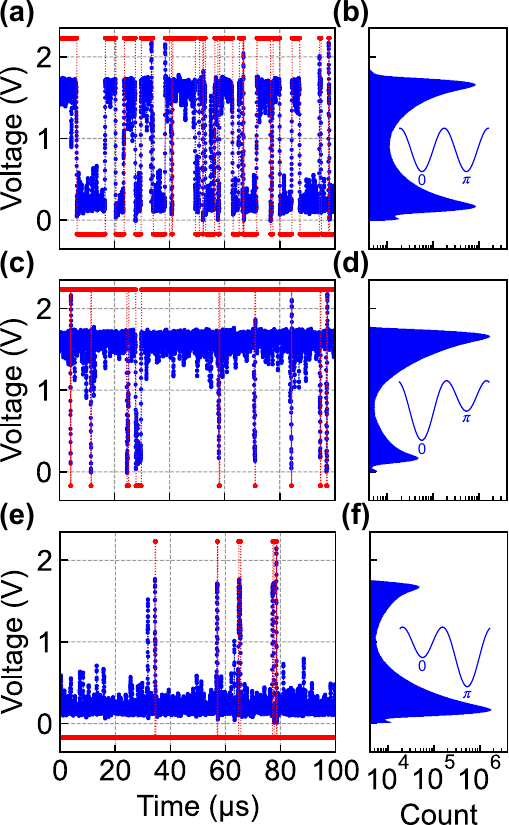}
  \caption{Typical voltage-time traces and corresponding count histograms measured at the output of the phase-readout circuit. (a) Voltage-time trace acquired in the absence of RF bias excitation, for a SHIL drive applied at $f_{\mathrm{IL}}=488$~MHz with power $P_{\mathrm{IL}}=-9$~dBm. The blue curve shows the measured trace, while the red points show the corresponding binarized signal after excluding incomplete transitions and local noise around the stable states. (b) Corresponding count histogram obtained from the full 500~ms-long voltage-time trace, containing approximately $2.4\times10^5$ transitions. The horizontal axis is logarithmic. The small peak near 1.7~V is a parasitic artifact arising from the phase comparator response during rapid transitions. (c),(e) Voltage-time traces measured with the RF bias excitation turned on at a power of $\mathrm{P}_\mathrm{bias}=-40$~dBm, for bias phases $\alpha=5\pi /18$~rad and $\alpha=23\pi /18$~rad, respectively. (d),(f) Corresponding histograms, showing that the RF bias excitation favors the $0$ state for $\alpha=5\pi /18$~rad and the $\pi$ state for $\alpha=23\pi /18$~rad.}
  \label{fig:3}
\end{figure}

Representative voltage-time traces measured at the output of the phase-readout circuit are shown in Fig.~\ref{fig:3}~(a), (c), and (e) for different conditions of the RF bias excitation. 
Each panel displays a 100~µs window extracted from a 500~ms-long voltage-time trace, containing overall between $8\times10^4$ and $2.4\times10^5$ transitions between two well-defined voltage levels. In the chosen readout configuration, these high- and low-voltage levels correspond to the $0$ and $\pi$ states, and are centered around about 1.7~V and 0.1~V, respectively. 
We therefore interpret the stochastic switching between these two voltage levels as random jumps of the STNO phase between the $0$ and $\pi$ states.

The voltage-time trace in Fig.~\ref{fig:3}~(a) was acquired with the RF bias excitation turned off. In this case, the corresponding count histogram in Fig.~\ref{fig:3}~(b) shows that the $0$ and $\pi$ states are populated with equal probability, consistent with the unbiased stochastic switching reported in Ref.~\cite{phan2024unbiased}.
When the RF bias excitation is turned on with an applied bias power of $\mathrm{P}_\mathrm{bias}=-40$~dBm, the voltage-time traces and their corresponding count histograms exhibit a pronounced bias between the two phase states.
In Fig.~\ref{fig:3}~(c) and (d), the bias phase is set to $\alpha=5\pi /18$~rad, whereas in Fig.~\ref{fig:3}~(e) and (f) the same bias power is used with $\alpha=23\pi/18$~rad. These two phases differ by $180^\circ$, which reverses the effect of the RF bias excitation on the occupation probabilities. As shown by the histograms, the $0$ state is favored for $\alpha=5\pi /18$~rad, while the $\pi$ state is favored for $\alpha=23\pi/18$~rad.

\subsection{Mean dwell time and probability extraction for the 0 and \texorpdfstring{\boldsymbol{$\pi$}}{pi}-phase states} \label{II.D}

To move beyond the first qualitative observations of Sec.~\ref{II.C}, we next quantify two metrics. The first is the mean dwell time, the average time spent in the $0$ ($\pi$) state before switching to the $\pi$ ($0$) state.
The second is the corresponding occupation probability of the $0$ ($\pi$) state. We determine how both depend on the bias phase $\alpha$ and on the applied power of the RF bias excitation.
To determine these quantities accurately, we use a numerical digitization procedure described in Appendix~\ref{A.C}. The procedure removes incomplete switching events and small fluctuations around the two stable output-voltage states of the phase-readout signal. The resulting refined digital signal, shown in red in Fig.~\ref{fig:3}~(a), (c), and (e), retains only the complete transitions between the $0$ and $\pi$ states. The measured analog dynamics is thus reduced to a discrete two-state stochastic process.

From this refined signal, we extract the duration of each residence in the $0$ and $\pi$ states over many subsequent phase-jumps, thereby obtaining two statistical sets of dwell times. The resulting dwell-time distributions exhibit an exponential-like form (see Appendix~\ref{A.C}), consistent with an approximate description of the measured dynamics as a discrete two-state Markov process~\cite{phan2024unbiased}. The cumulative distribution function of each set is then fitted with a modified exponential-like distribution that accounts for the finite bandwidth of the measurement setup. In each fit, the only free parameters are the mean dwell times, $\tau_0$ for the $0$ state and $\tau_{\pi}$ for the $\pi$ state. Throughout the rest of this work, $\tau_0$ and $\tau_{\pi}$ are used as key quantitative measures of the temporal stability of the two metastable phase states. For each voltage-time trace shown in Fig.~\ref{fig:3}, the extracted values of $\tau_{0}$ and $\tau_{\pi}$ are reported in Tab.~\ref{tab:A.C.1} of Appendix~\ref{A.C}.

Another crucial metric used in this work is the occupation probability of the $0$ ($\pi$) state. From the mean dwell times, we define the occupation probabilities of the $0$ ($\pi$) state as $P_{0(\pi)}=\tau_{0(\pi)}/(\tau_{0}+\tau_{\pi})$. This quantity measures the relative statistical occupancy of the two metastable phase states. For each voltage-time trace shown in Fig.~\ref{fig:3}~(a), (c) and (e), the corresponding $P_{0}$ is approximately $0.50$, $0.95$, and $0.05$, respectively (see Tab.~\ref{tab:A.C.1} in Appendix~\ref{A.C}). Thus, we conclude quantitatively that the bias phase $\alpha$ can be used to favor one metastable phase state over the other.

\section{Deterministic dwell time control of the 0 and \texorpdfstring{\boldsymbol{$\pi$}}{pi}-phase states} \label{III}

\begin{figure*} \centering
	\includegraphics[width=170mm]{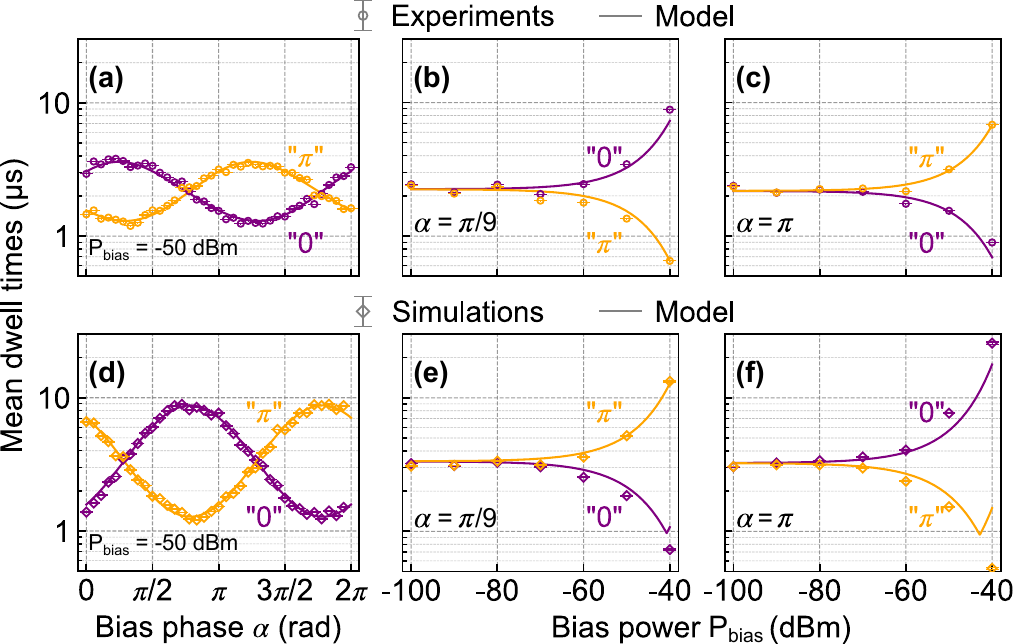}
    \caption{Mean dwell times of the 0 and $\pi$-phase states in the SHIL regime under RF bias excitation. (a) Mean dwell times as a function of the bias phase $\alpha$, varied from $0$ to $2\pi$~rad at fixed bias power $\mathrm{P}_\mathrm{bias}=-50$~dBm.
    (b) and (c) Mean dwell times as a function of the bias power $\mathrm{P}_\mathrm{bias}$, swept from $-100$~dBm to $-40$ dBm, for fixed bias phases $\alpha=\pi/9$ rad and $\alpha=\pi$ rad, respectively. Circular markers correspond to experimental values extracted using the method described in Appendix~\ref{A.C}, while the solid lines are analytical fits based on Eq.~\eqref{eq:dwell_times}, with details given in Appendix~\ref{A.G}. (d)--(f) show the corresponding numerical simulation results obtained from the finite-temperature Thiele equation approach described in Appendix~\ref{A.D}. Diamond symbols denote numerical values extracted from the simulations using the same procedure as for the experimental data. The analytical fits reproduce the numerical trends well, with noticeable deviations in (e) and (f) at the highest bias power.}
    
	\label{fig:4}
\end{figure*}
Beyond setting the occupation probabilities, the mean dwell times directly determine the timescale of the stochastic switching process and therefore the effective sampling rate at which a hardware $p$-bit or stochastic neuron can operate.
To investigate more systematically the control exerted by the RF bias excitation, we record phase-readout voltage-time traces for bias phases $\alpha$ ranging from $0$ to $2\pi$~rad in steps of $\pi/18$~rad, at a fixed bias power $\mathrm{P}_\mathrm{bias}=-50$~dBm. We then analyze these traces following the technique introduced in Sec.~\ref{II}~(C) to extract the mean dwell times $\tau_0$ and $\tau_{\pi}$ of the two metastable phase states. 

The resulting dwell times are summarized in Fig.~\ref{fig:4}~(a). As a function of $\alpha$, $\tau_{0}$ and $\tau_{\pi}$ exhibit opposite $2\pi$-periodic sinusoidal-like variations on the logarithmic scale, consistent with the periodic nature of the bias phase. This qualitatively indicates that the logarithm of the mean dwell times varies sinusoidally with the bias phase $\alpha$, i.e. $\log(\tau_{0,\pi})\propto \pm\sin(\alpha-\alpha_0)$, with an effective phase offset $\alpha_0 \approx \pi/13$~rad. The slight deviations from a fully symmetric behavior can likely be attributed to the finite number of recorded switching events, which limits the statistical precision of the extracted mean dwell times (see Appendix~\ref{A.C}).
More quantitatively, $\tau_{0}$ varies from approximately $1.24$~µs to $3.8$~µs, whereas $\tau_{\pi}$ ranges from approximately $1.19$~µs to $3.55$~µs. Relative to the mean dwell time $\tau_\mathrm{unbias}\approx 2.33$~µs obtained when the RF bias excitation is turned off, these values correspond to variations of about $-47\%$ to $+63\%$ for $\tau_{0}$ and $-49\%$ to $+52\%$ for $\tau_{\pi}$. 
Although the accessible tuning range of the mean dwell times remains limited to approximately $2.5$~µs, we show that varying the bias phase $\alpha$ enables continuous control over the favored metastable phase state.


Next, we show that this accessible tuning range can itself be adjusted through the bias power $\mathrm{P}_\mathrm{bias}$ by fixing the bias phase at $\alpha=\pi/9$~rad and sweeping $\mathrm{P}_\mathrm{bias}$ from $-100$~dBm to $-40$~dBm in steps of $10$~dBm. The resulting mean dwell times are summarized in Fig.~\ref{fig:4}~(b).
For bias powers below about $-70$~dBm, the difference between $\tau_{0}$ and $\tau_{\pi}$ remains smaller than $0.5$~µs, indicating no significant preference for either the $0$ or the $\pi$ state in this low-power regime. As the bias power increases, the contrast between the mean dwell times of the two metastable phase states becomes progressively stronger, with $\tau_{0}$ increasing and $\tau_{\pi}$ decreasing monotonically. At the highest applied bias power, $\mathrm{P}_\mathrm{bias}=-40$~dBm, this contrast becomes particularly pronounced, with mean dwell times of about $8.85$~µs for the $0$ state and $0.65$~µs for the $\pi$ state, corresponding to variations of about $+280\%$ and $-72\%$, respectively, relative to the unbiased case.

Changing the bias phase to $\alpha=\pi$~rad reverses the trend observed in Fig.~\ref{fig:4}~(b), as shown in Fig.~\ref{fig:4}~(c). In this case, $\tau_{0}$ decreases monotonically while $\tau_{\pi}$ increases monotonically, so that the $\pi$ state becomes progressively favored over the $0$ state. At $\mathrm{P}_\mathrm{bias}=-40$~dBm, the mean dwell times reach about $0.9$~µs for the $0$ state and $6.82$~µs for the $\pi$ state, corresponding to variations of about $-61\%$ and $+193\%$, respectively, relative to the unbiased case. As above, for bias powers below about $-70$~dBm, the difference between $\tau_{0}$ and $\tau_{\pi}$ remains small, indicating no significant preference for either state in this low-power regime. Overall, from these experimental observations, $\alpha$ and $\mathrm{P}_\mathrm{bias}$ appear as two complementary control parameters of the stochastic phase dynamics, with $\alpha$ determining which metastable phase state is favored and $\mathrm{P}_\mathrm{bias}$ controlling how strongly it is favored.

The above experimental observations are further supported by extensive numerical simulations based on the finite-temperature Thiele equation approach, detailed in Appendix~\ref{A.D}. Applying the same voltage-time trace analysis as for the experimental data (Sec.~\ref{II}~(C-D)), the simulations qualitatively reproduce the evolution of the mean dwell times of the 0  and $\pi$-phase states with bias phase $\alpha$ and bias power $\mathrm{P}_\mathrm{bias}$, as shown in Fig.~\ref{fig:4}~(d)--(f).
Some discrepancies nevertheless remain between experiment and simulation. The dwell-time curves versus $\alpha$ are phase-shifted, as shown in Fig.~\ref{fig:4}~(d), and the power-dependent trends at fixed nominal $\alpha$ are reversed in Fig.~\ref{fig:4}~(e) and (f), with different extrema. This likely reflects a different effective phase offset $\phi_0$, mentioned in Sec.~\ref{II}~(A), due in particular to uncompensated delays in the experimental setup, as well as differences in oscillation amplitude between the simulated and experimental free-running and SHIL regimes.

\section{Deterministic probability control of the 0 and \texorpdfstring{\boldsymbol{$\pi$}}{pi}-phase states} \label{IV}

\begin{figure*} \centering
    \includegraphics[width=170mm]{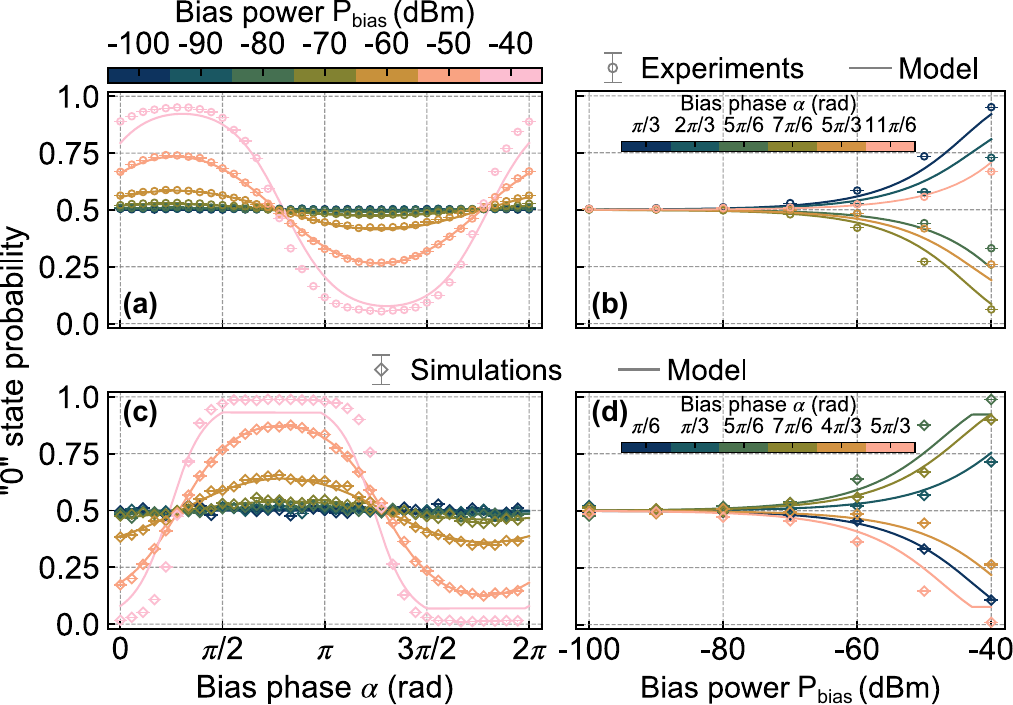}
    \caption{Occupation probabilities of the $0$-phase state in the SHIL regime under RF bias excitation. (a) $0$-phase state occupation probabilities as a function of the bias phase $\alpha$, varied from $0$ to $2\pi$~rad. (b) $0$-phase state occupation probabilities as a function of the bias power $P_\mathrm{bias}$, swept from $-100$~dBm to $-40$~dBm, for multiple $\alpha$ values listed in the colorbar. Circular markers correspond to experimental values computed from the mean dwell times using the formula $P_{0}=\tau_{0}/(\tau_{0}+\tau_{\pi})$. The solid lines are constructed from the analytical fits of the mean dwell times based on Eq.~\eqref{eq:dwell_times} (see Appendix~\ref{A.G} for details), using the same probability formula. (c) and (d) show the corresponding numerical simulation results obtained from the finite-temperature Thiele equation approach described in Appendix~\ref{A.D}. Diamond symbols denote numerical values extracted from the simulations using the same procedure as for the experimental data. The analytical fits reproduce the numerical trends well, with noticeable deviations at the highest bias power.}
    \label{fig:5}
\end{figure*}
Beyond the mean dwell times $\tau_0$ and $\tau_\pi$, the stochastic phase dynamics can also be described through the occupation probabilities $P_0$ and $P_\pi$. As introduced in Sec.~\ref{II}~(D), these probabilities are directly obtained from the mean dwell times. Fig.~\ref{fig:5}~(a) shows the extracted probability $P_0$ of the $0$-phase state as a function of the bias phase $\alpha$, for $\mathrm{P}_\mathrm{bias}$ ranging from $-100$ to $-40$~dBm in steps of $10$~dBm, as indicated by the colorbar.
Qualitatively, for $\mathrm{P}_\mathrm{bias}$ below about $-40$~dBm, $P_0$ follows a phase-shifted sinusoidal-like variation around $0.5$, i.e. $P_0 - 0.5 \propto \sin(\alpha-\alpha_0)$, with an effective phase offset $\alpha_0 \approx \pi/8$~rad.
At $\mathrm{P}_\mathrm{bias}=-40$~dBm, however, a slight departure from this simple sinusoidal-like behavior can be noticed, with the extrema becoming visibly flattened as $P_0$ asymptotically tends toward $0$ and $1$.
As $\mathrm{P}_\mathrm{bias}$ decreases, the probability variation progressively vanishes and $P_0 \approx 0.5$ over the full $\alpha$ range, thereby recovering the unbiased limit~\cite{phan2025leveraging}.

An alternative view is provided by Fig.~\ref{fig:5}~(b), which shows the extracted probability $P_0$ as a function of $\mathrm{P}_\mathrm{bias}$ for several fixed $\alpha$, as indicated by the colorbar. For each $\alpha$, $P_0$ evolves monotonically with increasing $\mathrm{P}_\mathrm{bias}$, starting near $0.5$ at low bias power and progressively moving away from the unbiased limit. Depending on $\alpha$, this evolution drives $P_0$ either above or below $0.5$, reflecting the preferential occupation of one of the two phase states. For the bias phase $\alpha$ producing the strongest probabilistic imbalance, $P_0$ approaches values close to $0$ or $1$ at high $\mathrm{P}_\mathrm{bias}$, with a slight bending of the curves that suggests the onset of saturation near these bounds.

The numerical simulations (see Appendix~\ref{A.D}) shown in Fig.~\ref{fig:5}~(c) and (d), reproduce the main experimental trends of the $0$ state probability as a function of both $\alpha$ and $\mathrm{P}_\mathrm{bias}$. They also reveal more clearly the onset, at $\mathrm{P}_\mathrm{bias}=-40$~dBm, of a regime where the RF bias excitation is no longer a weak perturbation of the SHIL dynamics but starts to compete with it. In this regime, the phase-state probabilities become extremely asymmetric, with the extrema becoming strongly flattened as $P_0$ asymptotically tends toward $0$ and $1$, indicating that one phase state becomes strongly favored while the other loses stability. In our interpretation, increasing $\mathrm{P}_\mathrm{bias}$ further would ultimately suppress the SHIL phase binarization and drive synchronization to the RF bias excitation alone. A more detailed investigation of this regime is left for future work.

Fig.~\ref{fig:5} shows that the RF bias excitation enables deterministic control of the phase-state probability, analogous to the tunable bias of a $p$-bit. The bias phase $\alpha$ selects the statistically favored phase state, while $\mathrm{P}_\mathrm{bias}$ sets the strength of the resulting occupation imbalance, from the unbiased limit $P_0 \approx 0.5$ to strongly asymmetric probabilities $P_0 \approx 0$ or $1$.

\section{Toy model} \label{V}

\begin{figure} \centering
	\includegraphics[width=85mm]{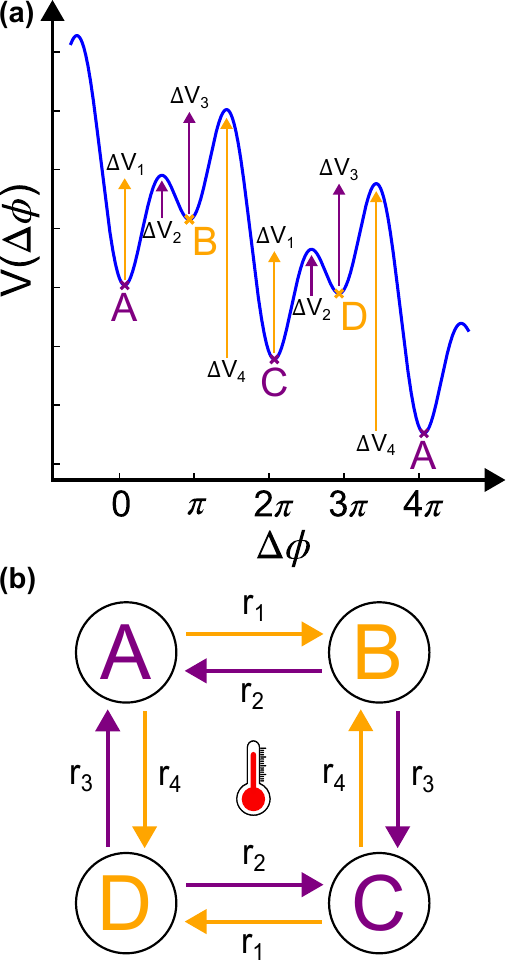}
	\caption{(a) Schematic quasipotential of an STNO in the SHIL regime under RF bias excitation, shown over a $4\pi$ interval of $\Delta\phi$ to display four successive extrema $A$--$D$ and the eight nearest-neighbor potential barriers between them. The states $A$ and $C$ correspond to the $0$ state, while $B$ and $D$ correspond to the $\pi$-state. (b) Corresponding four-state Markov model for the stochastic phase dynamics, where arrows denote thermally activated jumps between neighboring states with rates set by the associated barrier heights.}
	\label{fig:6}
\end{figure}
To gain physical insight into the experimental and numerical results presented in the previous sections, it is useful to describe the phase dynamics of an STNO in the SHIL regime under an additional RF bias excitation in terms of an effective phase quasipotential landscape (see Fig.~\ref{fig:1}~(b)), extending the model of Ref.~\cite{phan2024unbiased}. 
For the STNO subjected to both the SHIL drive and the RF bias excitation, the phase-rate equation defined in Appendix~\ref{A.E} reduces, under the condition $f_{\mathrm{bias}} = f_{\mathrm{IL}}/2$, to a modified Adler-like equation~\cite{adler1946, slavin_nonlinear_2009}. Integrating this equation yields a phase quasipotential of the phase difference $\Delta\phi$, written as
\begin{equation}
    \begin{aligned} 
        V(\Delta\phi) = & \dfrac{\delta}{2} \Delta\phi  - \dfrac{\Omega}{4}\cos\left(2\Delta\phi\right)
         - \Omega^\prime\cos\left(\Delta\phi - \alpha + \alpha_0\right),
    \end{aligned} \label{eq:potential}
\end{equation}
where $\delta = 2\pi(2f_{\mathrm{STNO}}-f_{\mathrm{IL}})$ is the frequency detuning between the SHIL drive and twice the free-running STNO frequency. The parameters $\Omega$ and $\Omega^\prime$ denote the effective synchronization bandwidths associated with the SHIL drive and the additional RF bias excitation, respectively. The bandwidth $\Omega$ is defined from the maximum detuning from $f_{\mathrm{IL}}$ for which SHIL is maintained, whereas $\Omega^\prime$ is defined from the corresponding locking bandwidth associated with the RF bias excitation and is therefore referenced to $f_{\mathrm{bias}}$.
$\alpha$ is the bias phase, and $\alpha_0$ is an effective constant phase offset arising from the STNO nonlinearity together with phase shifts occurring during the transmission of the RF bias excitation and of the 10~MHz reference.

The corresponding quasipotential is illustrated schematically in Fig.~\ref{fig:6}~(a). It can be viewed as a washboard potential whose corrugation results from the superposition of a $\pi$-periodic SHIL contribution and a $2\pi$-periodic contribution from the RF bias excitation. 
The washboard potential is tilted whenever the detuning $\delta$ is non-zero, the slope being set by its magnitude.
It therefore exhibits alternating local minima and maxima as a function of $\Delta\phi$. The minima labeled A and C correspond to the $0$ state, repeated modulo $2\pi$, while the minima labeled B and D correspond to the $\pi$ state. In the limit $\Omega^\prime=0$, the quasipotential reduces to the tilted $\pi$-periodic washboard potential of the SHIL oscillator studied in Ref.~\cite{phan2024unbiased, phan2025leveraging}.


We assume that phase jumps between adjacent potential minima are thermally activated and that the mean dwell times are much longer than the oscillation period, consistent with the quasi-stationary auto-oscillation regime~\cite{kim2008generation, slavin_nonlinear_2009}. Under these assumptions, the transition rates of phase jumps are expected to depend predominantly on the barrier heights, i.e., on the differences in the quasipotential between each minimum and the adjacent maxima.
Using the $2\pi$-periodic symmetry of the quasipotential, the eight possible transition barriers reduce to four distinct ones. For phase-jumps with increasing $\Delta\phi$, the barriers associated with the paths A$\rightarrow$B and C$\rightarrow$D are identical and are denoted $\Delta V_1$, while those associated with B$\rightarrow$C and D$\rightarrow$A are identical and are denoted $\Delta V_3$. For phase jumps with decreasing $\Delta\phi$, the barriers associated with B$\rightarrow$A and D$\rightarrow$C are identical and are denoted $\Delta V_2$, while those associated with C$\rightarrow$B and A$\rightarrow$D are identical and are denoted $\Delta V_4$. 
This quasipotential picture can therefore be mapped onto the four-state Markov model shown schematically in Fig.~\ref{fig:6}~(b), where A, B, C, and D denote successive extrema of the phase landscape. Restricting the dynamics to nearest-neighbor jumps, the $2\pi$-periodic symmetry reduces the description to four distinct average phase-jump rates $r_1$ to $r_4$, associated respectively with the barriers $\Delta V_1$ to $\Delta V_4$.

Treating the RF bias excitation as a weak perturbation of the SHIL drive, i.e. $\epsilon = \Omega^\prime/\Omega\ll1$, first-order perturbation theory~\cite{nayfeh2011introduction, strogatz2018nonlinear} can be applied to Eq.~\eqref{eq:potential} to derive approximate analytical expressions for the stable and unstable phase states and for the associated potential barriers, as detailed in Appendix~\ref{A.E}. To first order in $\epsilon$, the four barrier heights read


\begin{equation}
    \begin{aligned}
        \Delta V_{1(3)} \approx \Delta V_{1(3)}^{\mathrm{0}}
        \mp 2\Omega^\prime\sin\left(\dfrac{\theta}{2} + \dfrac{\pi}{4}\right) \sin\left(\alpha-\alpha_0 - \dfrac{\pi}{4}\right), \\
        \Delta V_{2(4)} \approx \Delta V_{2(4)}^{\mathrm{0}}
        \pm 2\Omega^\prime\sin\left(\dfrac{\theta}{2} - \dfrac{\pi}{4}\right) \sin\left(\alpha-\alpha_0 + \dfrac{\pi}{4}\right). 
    \end{aligned} \label{eq:barriers}
\end{equation}
Here, $\Delta V_{1(3)}^{0}$ and $\Delta V_{2(4)}^{0}$ are the corresponding barrier heights in the absence of RF bias excitation, with $\Delta V_{1(3)}^{0} = (\theta/2+\pi/4)\delta + (\Omega/2)\cos(\theta)$ and $\Delta V_{2(4)}^{0} = (\theta/2-\pi/4)\delta + (\Omega/2)\cos(\theta)$, where $\theta=\arcsin(\delta/\Omega)$. The terms proportional to $\Omega^\prime$ represent the first-order correction induced by the RF bias excitation. In the case $\Omega^\prime=0$, Eq.~\eqref{eq:barriers} recovers the unbiased barrier heights, consistent with the corresponding expression reported in Ref.~\cite{phan2024unbiased, phan2025leveraging}.
In the quasi-stationary thermally activated regime, assuming that the system is at equilibrium with the thermal bath and neglecting the local curvature of the phase quasipotential, the average phase-jump rates $r_i$ associated with overcoming the barriers $\Delta V_i$ in the presence of thermal energy $k_BT$, where $k_B$ is the Boltzmann constant and $T$ the temperature, are expected to follow Arrhenius-type laws,
\begin{equation}
    r_i \approx r_0 \exp\left(-\dfrac{\Delta V_i}{\eta k_B T}\right),
    \label{eq:rate}
\end{equation}
where $\eta$ is a corrective scaling factor linking the amplitude of the phase fluctuations to the thermal energy. Here $\eta$ depends on the damping and on the STNO oscillation amplitude and can be estimated at the order-of-magnitude level from the STNO frequency linewidth~\cite{grimaldi2014response}. $r_0$ can be seen as an effective attempt rate for thermally activated phase jumps and is therefore expected, in our interpretation, not to exceed the oscillator's power relaxation rate~\cite{slavin_nonlinear_2009, grimaldi2014response}.

Within the four-state Markov description, analytical expressions for the mean dwell times in the $0$ and $\pi$ states can be readily derived, as detailed in Appendix~\ref{A.F}. These two states correspond respectively to the pairs of Markov states (A,C) and (B,D) shown in Fig.~\ref{fig:6}~(b). The model, therefore, directly relates the average phase-jump rates to the mean dwell times, which are given by
\begin{equation}
    \begin{aligned}
        \tau_{0} = \dfrac{1}{r_{1}+r_{4}}, 
        \qquad
        \tau_{\pi} = \dfrac{1}{r_{2}+r_{3}}.
    \end{aligned}
    \label{eq:dwell_times}
\end{equation}
Inserting Eq.~\eqref{eq:barriers} into Eq.~\eqref{eq:rate} used to evaluate Eq.~\eqref{eq:dwell_times} analytically, we reproduce with good agreement the experimental and numerical evolutions of the mean dwell times shown in Fig.~\ref{fig:4}. The corresponding fits are shown as solid lines, and all parameters used to fit the experimental and numerical results are reported in Tables~\ref{tab:A.G.1},~\ref{tab:A.G.2}, and~\ref{tab:A.G.3} in Appendix~\ref{A.G}. The fitted parameters of the model remain physically consistent. In particular, the extracted effective rates $r_0$ and scaling factors $\eta$ are of the same order as those expected from the oscillator dynamics, while the fitted values of $\Omega^\prime$ and $\alpha$ remain comparable to the experimental and numerical conditions, supporting the model's ability to predict the control parameters required to achieve a target mean dwell-time.

This model also clarifies the origin of the observed trends. The nearly sinusoidal dependence of $\log(\tau_{0,\pi})$ on $\alpha$ arises from the sinusoidal first-order corrections to the barrier heights in Eq.~\eqref{eq:barriers}. Likewise, the progressive separation of $\tau_0$ and $\tau_\pi$ with increasing $\mathrm{P}_\mathrm{bias}$ stems from the explicit dependence of the barriers on $\Omega^\prime$. 
Nevertheless, the agreement degrades at the highest bias power, $\mathrm{P}_\mathrm{bias}=-40$~dBm, where the RF bias excitation is no longer strictly a weak perturbation of the SHIL drive. In this regime, first-order perturbation theory is likely no longer sufficient to capture the barrier heights quantitatively.

Since the occupation probabilities are directly derived from the mean dwell times (see Sec.~\ref{II.D}), the quality of the model can be further assessed by comparing its predictions with the probability data of Fig.~\ref{fig:5}. As expected from the good agreement obtained for the dwell times, the model captures well the sinusoidal-like variation of the probability $P_0$ with $\alpha$ and its monotonic departure from the unbiased limit $P_0 \approx 0.5$ with increasing $\mathrm{P}_\mathrm{bias}$, for both experiments and numerical simulations (solid lines in Fig.~\ref{fig:5}~(a)--(d)). Nevertheless, deviations appear at the highest bias power $\mathrm{P}_\mathrm{bias} = -40$~dBm, where the model does not fully reproduce the saturation of $P_0$ near 0 and 1 visible in the $\alpha$-dependent curves of Fig.~\ref{fig:5}~(a) and~(c), nor the flattening onset in the power-dependent curves of Fig.~\ref{fig:5}~(b) and~(d). These deviations mirror those already identified for the dwell times and confirm that the first-order perturbative treatment of the barrier heights becomes insufficient when the RF bias excitation is no longer a weak perturbation of the SHIL drive.
We note that this limitation can be circumvented by evaluating the barrier heights numerically from the full quasipotential of Eq.~\eqref{eq:potential}, without applying perturbation theory, at the cost of losing closed-form analytical expressions. 
The deviations visible in Fig.~\ref{fig:4} and \ref{fig:5} at the highest $\mathrm{P}_\mathrm{bias}$ between the analytical model and the experimental and simulation data are therefore attributed to the first-order expansion of the barrier heights. 

\section{Discussion and conclusion} \label{VI}

We have experimentally demonstrated that the stochastic phase dynamics of an second-harmonic injection-locked STNO can be deterministically controlled by a weak RF bias excitation applied at the oscillator frequency. In the SHIL regime, the oscillator phase fluctuates between two degenerate attractors separated by $\pi$, and a weak RF bias excitation lifts this degeneracy. The bias phase $\alpha$ selects which phase state is statistically favored, while the bias power $\mathrm{P}_\mathrm{bias}$ sets the strength of the resulting occupation imbalance. This enables continuous tuning of the output probability from the unbiased limit $P_0 \approx 0.5$ to strongly asymmetric values approaching 0 or 1. A phenomenological model combining a washboard quasipotential with Arrhenius-type thermally activated escape rates quantitatively accounts for the observed trends across both experiments and numerical simulations, and provides predictive capability over the relevant parameter range.

A central motivation for this work is the growing interest in STNO-based Ising machines~\cite{albertsson2021ultrafast, raimondo2025high, raimondo2025adaptive, abderrazakSPIE}, where networks of coupled oscillators are used to solve combinatorial optimization problems by mapping oscillator phase onto spin variables. 
In such architectures, SHIL enables each oscillator to encode a binary Ising spin through binarized phase state of the oscillator. Each oscillator $i$ experiences an effective local field that combines the coupling contributions from neighboring oscillators $j$, which encode the interaction terms $J_{ij}$ of the Ising Hamiltonian, and an external bias, which encodes the field terms $h_i$. The probability control demonstrated here provides precisely this external bias capability, enabling each STNO, under SHIL, to function as a fully tunable Ising spin programmed through the amplitude and phase of an RF bias excitation.

The interaction terms $J_{ij}$ modify the probability of joint phase-states through the same mechanism as the RF bias excitation used in this work, although a network-level treatment lies beyond the scope of the present single-device study. With all oscillators of the network commonly under SHIL to the same drive, the signal received by oscillator $i$ from the oscillators $j$ coupled to it at time $t$ would take the form $v_i(t) \propto \sin(\pi f_\mathrm{IL}t + \alpha_i(t))$, where $\alpha_i(t)$ depends on the instantaneous phase configuration of the network. In the zero temperature limit, the network settles into a stable phase configuration, $\alpha_i$ becomes constant, and the coupling signal reduces to a pure sinusoidal drive formally equivalent to the RF bias excitation studied here. 


Away from this deterministic limit, the results presented here establish the injection-locked STNO as a nanoscale stochastic element that can serve as the building block of a probabilistic bit ($p$-bit)~\cite{faria2017low, sutton2017intrinsic, borders2019integer}, a fundamental primitive not only for Ising machines but also for invertible logic~\cite{camsari2017stochastic} and probabilistic inference circuits~\cite{sutton2017intrinsic}. However, the RF generation, routing, synchronization, and phase readout remain integration challenges to be addressed for full $p$-bit-level operation. In existing implementations based on superparamagnetic tunnel junctions~\cite{borders2019integer, singh2024cmos, iftakher2026intrinsic}, the $p$-bit probability is tuned through a DC input current that modulates the energy barrier between two magnetization states. The STNO-based approach demonstrated here offers a complementary route in which the stochastic degree of freedom is the oscillator phase, controlled through RF signals rather than DC currents. As for existing hardware implementations of $p$-bits at the nanoscale, a single STNO device can simultaneously provide both the stochastic fluctuation and the nonlinear response, without requiring an external noise source or a separate amplification stage. Importantly, because the phase-jump dynamics is governed by the oscillator frequency and relaxation rate, STNO-based $p$-bits carry the promise of accessing intrinsically fast stochastic dynamics, potentially reaching the gigahertz range in high-frequency oscillator geometries~\cite{litvinenko2022ultrafast}, thereby complementing existing magnetic tunnel junction approaches whose fluctuation rates are set by the thermal activation over magnetization energy barriers.

An additional opportunity specific to phase-based encoding is the ability to move beyond binary spin variables. Without the SHIL drive, the oscillator phase remains continuous and naturally maps onto an XY spin, as exploited in polariton and photonic XY-Hamiltonian solvers~\cite{berloff2017realizing, honari2020optical, chalupnik2024nanophotonic}. In a closely related direction, STNO arrays have been theoretically explored as complex-valued Hopfield networks, where both the oscillator phase and amplitude encode information beyond a pure XY description~\cite{prasad2022associative}. With injection-locking at a fractional harmonic $f_\mathrm{IL}/n$, the phase is instead confined to $n$ discrete stable states~\cite{LebrunPRL, Urazhdin2010Fractional}, enabling the native implementation of $n$ state Potts optimization models~\cite{mallick2022OPM, nikonov2023polychronous, lin2025OPM}. Above regimes lie outside the reach of strictly binary $p$-bits, even though the present work focuses on the Ising case $n=2$. Realizing this broader potential will require addressing the same integration challenges as for the binary $p$-bit operation discussed above, with the additional constraint of precise control over the relative phases of RF sources, an active direction of research toward CMOS-compatible spintronic-RF architectures~\cite{ma2021microwave}.

Furthermore, the $2\pi$-periodic sinusoidal-like dependence of $P_0$ on $\alpha$, with an amplitude controlled by $\mathrm{P}_\mathrm{bias}$, shown in Fig.~\ref{fig:5}, natively implements a tunable probabilistic activation function that can be used to emulate a stochastic binary neuron~\cite{mizrahi2018neural}. This is a key primitive in brain-inspired computing schemes that rely on stochastic encodings to perform probabilistic inference and sampling~\cite{faria2018implementing, daniels2020energy}. In accordance with Fig.~\ref{fig:5}(b,d), a monotonic activation spanning the full range from 0 to 1 can also be constructed by sweeping the bias amplitude at fixed $\alpha$ and shifting $\alpha$ by $\pm\pi$ when $P_0$ crosses the 0.5 level, thereby circumventing the periodicity constraint. This versatility makes the scheme relevant both for Boltzmann machines and belief networks~\cite{kaiser2022hardware}.
The quasipotential framework further suggests a possible route toward hardware-level annealing protocols~\cite{iftakher2026intrinsic} in STNO-based implementations. For instance, dynamically reducing $\mathrm{P}_\mathrm{bias}$ during operation progressively restores the barrier symmetry, which could be exploited in optimization schemes where the bias encodes a constraint that is gradually relaxed. A full exploration of such protocols and their applicability is, however, beyond the scope of the present work.

The present work focuses on a single vortex-based STNO operating in the sub-GHz range, chosen for its large microwave emission power and well-characterized dynamics. The underlying mechanism, however, is general and extends to any nonlinear oscillator exhibiting injection-locking and thermally activated phase jumps.
More generally, oscillator-based computing architectures extend well beyond the Ising machine paradigm and encompass a broad class of oscillatory neural networks in which the oscillator phases and their mutual synchronization are exploited for classification, pattern recognition, and associative memory tasks~\cite{csaba2020coupled, todri2024computing}. The probability control demonstrated here, which allows programming the preferred phase state of a single oscillator, constitutes a building block that is relevant to all such architectures.
Scaling to networks of coupled oscillators, where each node acts as a controllable artificial spin, $p$-bit, and stochastic neuron, remains an important challenge that will require addressing device-to-device variability. In this regard, the injection-locking scheme inherently mitigates frequency dispersion by forcing each oscillator to operate at a common reference frequency, as previously noted for unbiased operation~\cite{phan2024unbiased}. We also note that the current implementation relies on an analog phase-readout circuit and two RF sources, which adds scaling complexity compared to the resistive readout of superparamagnetic tunnel junctions~\cite{borders2019integer}. Integrating the RF generation and phase detection into CMOS-compatible circuitry~\cite{ma2021microwave} will be an important step toward scalability. Finally, the regime where the RF bias excitation competes with the SHIL drive rather than merely perturbing it was only briefly explored here, but revealed extreme probability asymmetries. A systematic investigation of this strongly driven regime may uncover functionalities beyond the perturbative framework developed in this work.

In summary, the interplay between injection-locking, thermal fluctuations, and a weak RF bias excitation transforms a single STNO into a deterministically programmable stochastic oscillator. The ability to encode, control, and read out probability using the phase of a nonlinear oscillator provides a versatile hardware primitive that bridges the gap between the rich physics of noisy spintronic oscillators and the growing demand for tunable stochastic elements in Ising machines, probabilistic computing, and brain-inspired architectures.

\section*{Acknowledgements}
This work was funded by the program CEA FOCUS “Numérique Frugal”, and supported by ANR via grant SpinIM project ANR-22-CE24-0004, NSF-ANR via grant StochNet Project ANR-21-CE94-0002-01 and NSF grant number CCF-CISE-ANR-FET-2121957, and partially supported by Grenoble INP Bourse Présidence and MIAI@Grenoble Alpes (ANR-19-P3IA-0). The authors would like to thank Eyub Yildiz, and Clément De Barbarin for fruitful discussions, technical support, and suggestions.

\FloatBarrier
\appendix
\label{appendix}

\section{Vortex-Based Spin-Torque Nano-Oscillator and Field line Structures} \label{A.A}
We use vortex-based STNOs (V-STNOs) developed at the International Iberian Nanotechnology Laboratory (INL). The magnetic tunnel junctions were fabricated from a layer stack comprising Ta(5)/CuN(50)/Ta(5)/CuN(50)/Ta(5)/Ru(5)/IrMn(6) \\
/CoFe\textsubscript{30}(2.6)/Ru(0.85)/CoFe\textsubscript{40}B\textsubscript{20}(1.8)/MgO/VFL \\
/Ta(10)/CuN(30)/Ru(7), where the numbers in parentheses indicate thicknesses in nanometers. The materials for the magnetic tunnel junction were deposited via a sputtering process, and the devices were patterned using ion beam milling and optical lithography at INL. The vortex-state free layer (VFL) consists of CoFe\textsubscript{40}B\textsubscript{20}(2.0)/Ta(0.2)/NiFe(7). The experiments conducted involved devices with a nominal diameter of 370 nm. Under the magnetic field conditions used in this work, the static resistance of the device in the vortex state was approximately 44.76 $\Omega$, featuring a tunneling magnetoresistance of about 43.74\% and a resistance-area product nearing 3.75~$\Omega\cdot$\textmu m\textsuperscript{2}. 
The field line is made of a 300~nm thick AlSiCu, 3~\textmu m wide and 39~\textmu m long, separated from the devices' top electrode with a 300~nm Al\textsubscript{2}O\textsubscript{3} layer for full electric isolation. The field line resistance is around 6~$\Omega$, with a linear current-to-field conversion factor of approximately $\sigma=0.18$~mT/mA.

\section{Free-Running and Injection-Locked Phase Dynamics of the Spin-Torque Nano-Oscillator} \label{A.B}
\begin{figure} \centering
	\includegraphics[width=85mm]{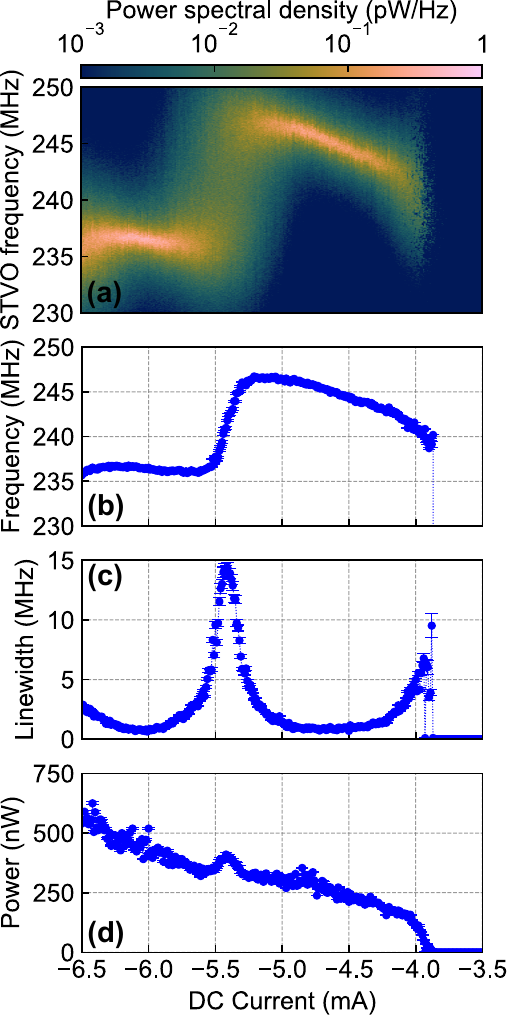}
	\caption{(a) Experimental power spectral density measured for the studied STNO in free-running regime corresponding to the absence of an external microwave drive at a fixed perpendicular applied magnetic field $\mu_0H_{\mathrm{DC}}\approx81.29$~mT. The applied current is swept on the x-axis from -3.5~mA to -6.5~mA. The bright regions illustrate the evolution of the free-running frequency as a function of the applied DC current. Two dynamical regimes can be distinguished in the free-running spectra. In our interpretation, the low-current regime, extending approximately from $-6.5$ to $-5.5~\mathrm{mA}$, corresponds to the vortex gyrotropic mode, whereas the high-current regime, above approximately $-5.5~\mathrm{mA}$, corresponds to a dynamical C-state mode. b) Evolution of the peak free-running frequency versus applied DC current. The frequency is evaluated from the Lorentzian fit of experimental frequency spectra. (c) Corresponding evolution of the linewidth extracted from Lorentzian fits. (d) Corresponding integrated power evaluated from the fits.}
    \label{fig:A.1}
\end{figure}

Fig.~\ref{fig:A.1} depicts the DC current dependence of the spectral characteristics of the free-running V-STNO device, in the $-6.5$~mA to $-3.5$~mA current range, for a mostly out-of-plane applied DC magnetic field of $\mu_0H_{\mathrm{DC}}\approx81.29$~mT. 
Fig.~\ref{fig:A.1}~(a) shows a colormap of the raw power spectral density measured in the free-running regime of the V-STNO. The spectrum exhibits a more complex behavior than expected for an ideal gyrotropic oscillator described solely by the Thiele model, with two distinct dynamical regimes over the explored current range. In our interpretation, the regime for currents below $-5.5~\mathrm{mA}$ corresponds to the vortex gyrotropic mode, whereas the regime for currents above $-5.5~\mathrm{mA}$ corresponds to a dynamical C-state mode~\cite{Wittrock2021C-state}.

Fig.~\ref{fig:A.1}~(b),~(c)~and~(d) show, respectively, the peak frequency, linewidth (full width at half maximum), and integrated power extracted from Lorentzian fits applied to the smoothed, instrument-baseline-subtracted spectra.
A transmission-line loss of approximately 10~dB was measured between the oscillator and the spectrum analyzer, as illustrated in Fig.~\ref{fig:2}. The power values reported for the V-STNO emission in Fig.~\ref{fig:A.1}~(d) were corrected for this loss, by adding 10 dB to the spectrum-analyzer readings.
The used operating point, detailed in Sec.~\ref{II.B}, was chosen to maximize synchronization capability of the device to external signals.

\begin{figure} \centering
	\includegraphics[width=85mm]{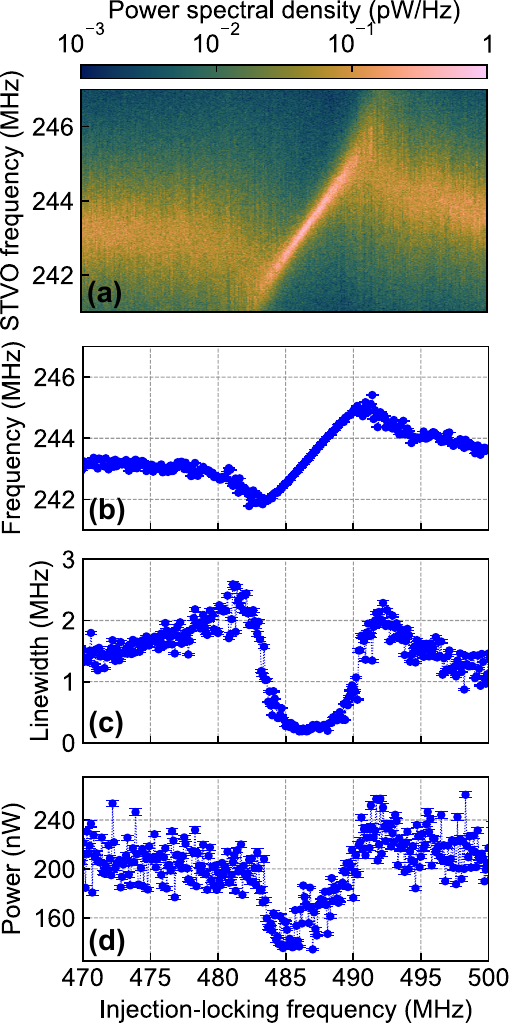}
	\caption{(a) Power spectral density corresponding to the microwave emissions of the studied STNO obtained at -4.4~mA DC applied current and in the presence of an external microwave signal with a frequency swept from 470~MHz to 500~MHz. The injection-locking power is maintained constant at -9~dBm. (b) Evolution of the peak frequency of the STNO from Lorentzian fits versus external drive frequency. (c) Corresponding evolution of the extracted linewidth versus injection-locking frequency exhibiting an increase and reduction of the linewidth, respectively, at the borders and center of the injection-locking range. (d) Evolution of integrated power evaluated from the fits versus the external drive frequency.}
    \label{fig:A.2}
\end{figure}

To explore the second-harmonic injection-locking behavior ($f_{\mathrm{IL}}\approx2f_0$), experiments were conducted at the established operating point detailed in Sec.~\ref{II.B}, where a constant current was maintained while varying the frequency of the external RF voltage drive. The power of the RF voltage source was set to $\mathrm{P}_{\mathrm{IL}}=$-9~dBm, but an additional $\approx-9.5$~dB attenuation from the used transmission line is to be taken into account. The V-STNO therefore receives approximately -18.5~dBm at its input. Fig.~\ref{fig:A.2}~(a) illustrates the raw power spectral density of the measured V-STNO, demonstrating how the oscillator frequency evolves as a function of the injection-locking frequency. A distinct locking region can be identified where the oscillator synchronizes with the applied RF voltage. This behavior confirms second-harmonic injection-locking, where the external drive influences the oscillator's frequency. Fig.~\ref{fig:A.2}~(b),~(c)~and~(d) show respectively the fitted frequency, linewidth, and integrated power of the measured V-STNO spectral emission as a function of the injection-locking frequency, using the same analysis described in the free-running case. Table~\ref{tab:A.B.1} contains the fitted spectral characteristics, in the free-running and SHIL regimes.
We observe an improvement of the oscillation coherence within the locking range from the linewidth reduction~\cite{LebrunPRL}, as well as the decrease of the integrated power. It should be noticed that the transmission line connecting the field line of the V-STNO and the bias RF source adds approximately $-2.5$~dB of attenuation. 


\begin{table} \centering
    \begin{tabular}{cc|c|c} 
          \textbf{Property} & \textbf{Unit} & \textbf{Free-running} & \textbf{SHIL} \\ \hline
        
        Frequency & MHz & 243.76 & 243.94 \\
        
        Linewidth & kHz & 1064.88 & 279.11 \\
        
        Integrated Power & nW & 225.5 & 151.57  \\ \hline
    \end{tabular}
    \caption{Results of the Lorentzian peak fitting of the V-STNO's microwave generation in the free-running and injection-locked at the second harmonic regimes. The fitting errors are at least two orders of magnitude smaller than the fitting parameters, and are therefore omitted.}
    \label{tab:A.B.1}
\end{table}

\section{Evaluation of Phase-State Dwell Times} \label{A.C}
 To access the mean dwell times of the stable phase states of a second-harmonic injection-locked STNO, in line with Ref.~\cite{phan2024unbiased}, we developed a measurement and analysis protocol of the measured time traces of the real-time stochastic jumps between these stable states.
First, we convert the analog signal recorded by the oscilloscope into a digital format, employing two thresholds for each phase state voltage as reference points. This method is crucial, as it accurately separates full phase jumps from "incomplete" jumps called excursions. 
\begin{figure} \centering
	\includegraphics[width=85mm]{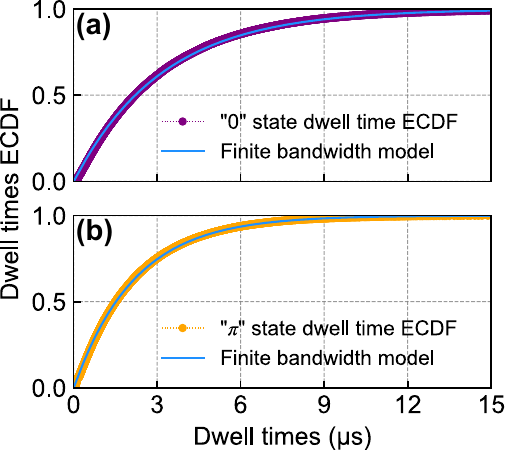}
    \caption{Empirical cumulative distribution function (ECDF) of the extracted dwell times for the 0 (a) and $\pi$ (b) phase states, obtained from the oscilloscope time traces at a bias power of $-40$~dBm and bias phase $\alpha = 13\pi/18$~rad. The blue curve shows the corresponding pseudo-exponential distribution CDF modified by the finite bandwidth of the measurement process (see Eq.~\eqref{eq:A.C.1}), used to fit the empirical data.}
  \label{fig:A.3}
\end{figure}
\FloatBarrier
The definition of a complete phase jump can vary. 
Here, we define it from the analog phase-readout voltage as a transition between the two voltage levels associated with the $0$ and $\pi$ stable states. To robustly distinguish these two levels, we introduce two voltage thresholds, approximately corresponding to the boundaries of the intermediate region separating the two stable states. A complete jump is then counted only when the signal successively crosses both thresholds, indicating a transition from one stable state to the other.
We chose to determine these thresholds from the inflection points of the probability distribution function obtained from the full voltage time trace. These points approximately correspond to the voltage boundaries separating the two stable levels from the intermediate transition region. Practically, they were identified as the voltage values at which the derivative of the histogram with respect to voltage reaches an extremum.

After digitization, we pinpointed the timestamps of each phase jump, enabling us to calculate the dwell time preceding each jump. We then analyzed the statistical distribution of these dwell times by examining the empirical cumulative distribution function (ECDF) for both the 0 and $\pi$-states, respectively shown in Fig.~\ref{fig:A.3}~(a)~and~(b). The ECDF closely resembles an exponential-like distribution, modified by the finite bandwidth of the measurement apparatus, which becomes important at shorter time scales. The effects of this finite bandwidth can be taken into account by employing the four-state Markov approach developed in~\cite{naaman2006poisson}. According to this method, instead of fitting to an exponential distribution, the ECDF is represented by a different distribution corresponding to the 0-state dwell times, following the formula
\begin{equation}
    \begin{aligned}
    F_A(t; \Gamma_0, \Gamma_\pi, \gamma) = & 1 - \dfrac{1}{2}\Biggl(\dfrac{G_1 + G_2}{G_2}\cdot\exp\left(\dfrac{G_2 - G_1}{2}t\right) \\
    & + \dfrac{G_2 - G_1}{G_2}\cdot\exp\left(-\dfrac{G_1 + G_2}{2}t\right)\Biggr),
    \end{aligned} \label{eq:A.C.1}
\end{equation}
where
\begin{equation}
    \begin{aligned}
        G_1= \Gamma_0 + \Gamma_\pi + \gamma, 
        \qquad
        G_2 = \dfrac{1}{2}\left(G^2_1 - 4\gamma\Gamma_0\right). 
    \end{aligned} \label{eq:A.C.2}
\end{equation}

In this more intricate ECDF, the parameter $\gamma$ represents the sampling rate that corresponds to the physical bandwidth of the measurement apparatus, whereas $\Gamma_0$ and $\Gamma_\pi$ denote the inverse of the mean dwell times for the 0 and $\pi$-states, respectively. To compute the ECDF for the dwell times in the $\pi$-state, one can simply swap $\Gamma_0$ with $\Gamma_\pi$. In the scenario of infinite measurement bandwidth $\gamma\to+\infty$, we recover the exponential distribution expressed as $F_0(t) = 1 - \exp(-\Gamma_{0}t)$.


By employing Eq.~\eqref{eq:A.C.1} and jointly fitting the ECDFs of the dwell times in the 0'' and $\pi$'' phase states, we obtained good agreement between the measured and fitted ECDFs, enabling the extraction of the mean dwell times of both phase states, as depicted in Fig.~\ref{fig:A.3}.
The finite bandwidth of our measurement circuit was fixed at $\gamma=100$~MHz. Table~\ref{tab:A.C.1} summarizes the mean dwell times computed using this method, for the measured time traces depicted in Fig.~\ref{fig:3}.
\begin{table} \centering
    \begin{tabular}{c|c|c|c} 
          \textbf{Fig.~\ref{fig:3}} & $\tau_{0}$ & $\tau_{\pi}$ & $P_{0}$ \\ \hline
        
        (a) & 2.33 $\pm$ $0.0005$ ~\textmu s & 2.31 $\pm$ $0.0005$ ~\textmu s & 0.5 $\pm$ $0.0001$ \\
        
        (b) & 10.63 $\pm$ $0.012$ ~\textmu s & 0.55 $\pm$ $0.0006$ ~\textmu s & 0.95 $\pm$ 0.0015 \\
        
        (c) & 0.59 $\pm$ $0.0009$ ~\textmu s & 10.35 $\pm$ $0.017$ ~\textmu s & 0.05 $\pm$ $0.0001$  \\ \hline
    \end{tabular}
    \caption{Extracted mean dwell times $\tau_0$ ($\tau_{\pi}$) of the 0 ($\pi$) state, and 0-state occupation probability for the measurements depicted in Fig.~\ref{fig:3}.}
    \label{tab:A.C.1}
\end{table}

\section{Numerical simulations using the stochastic Thiele equation approach}
\label{A.D}

To reproduce the experimental results, which were obtained in the gyrotropic-mode regime (see Appendix~\ref{A.B}), we restrict the modeling of the V-STNO dynamics to the vortex gyrotropic mode.
The gyrotropic mode corresponds to the quasi-circular orbital motion of the magnetic vortex core in the plane of the free layer. Neglecting vortex distortions, the free-layer magnetization dynamics is fully parameterized by the vortex core position $\mathbf{X}=(x,y)$, defined in Cartesian coordinates, whose time-evolution is governed by a corresponding Thiele equation~\cite{thiele_steady-state_1973, khalsa2015critical}, expressed as
\begin{equation}
    G\left(\mathbf{e_z}\times\dot{\mathbf{X}}\right) + D\dot{\mathbf{X}} -  k\mathbf{X} - \mathbf{F}_\mathrm{ST} + \mu\left(\mathbf{e}_z\times\mu_0\mathbf{H}_{x}\right) + \mathbf{f}(t)= 0,
    \label{eq:TEA}
\end{equation}
where $G$ is the gyrovector amplitude, $D$ the damping coefficient, $k$ the confinement stiffness of the vortex core, $\mathbf{F}_\mathrm{ST}$ the spin-transfer force, $\mu$ the magnetic coupling coefficient of the vortex core to the in-plane external field $\mathbf{H}_x$, and $\mathbf{f}(t)$ is the stochastic thermal fluctuation force at time $t$.
Thermal fluctuations are modeled as $\delta$-correlated Gaussian white noise satisfying
\begin{equation}
    \langle f_i \rangle = 0, \qquad \langle f_i(t)\,f_j(t') \rangle = \Gamma\,\delta_{ij}\,\delta(t - t'),
\end{equation}
where $\Gamma = 2 D k_B T / G^2 R_0^2$ sets the fluctuation amplitude through the coupling to the thermal bath at temperature $T$, following the fluctuation-dissipation relation~\cite{khalsa2015critical, kim2008generation, slavin_nonlinear_2009}, and $R_0$ is the free-layer radius.
The confinement stiffness $k$ takes into account the magnetostatic energy contribution $k_{\mathrm{ms}}$, and the Ampère-Oersted field contribution $\kappa_\mathrm{Oe}$~\cite{AbreuAraujo2022}, and is expressed as
\begin{align}
    k(s) =& k_{\mathrm{ms}}(s) + CJ\kappa_\mathrm{Oe}(s),
\end{align}
where $C=\pm1$ is the chirality of the vortex, and $J$ is the total out-of-plane electrical current density flowing through the free layer. The current density $J = J_{\mathrm{DC}} + J_{\mathrm{AC}}(t)$ comprises a DC component $J_{\mathrm{DC}}$ and a time-varying SHIL drive $J_{\mathrm{AC}}(t) = J_{\mathrm{IL}} \cos(2\pi f_{\mathrm{IL}}^{\mathrm{sim}} t)$.
Both $k_{\mathrm{ms}}(s)$ and $\kappa_\mathrm{Oe}(s)$ contributions depend on the radial vortex core displacement $s = (x^2 + y^2)^{1/2}$ and are expressed as even-order polynomial expansions,
\begin{align}
    k_\mathrm{ms}(s) =& k_{\mathrm{ms},0}\left(1+\zeta_\mathrm{ms} s^2 +\zeta_\mathrm{ms}^\prime s^4 +\zeta_\mathrm{ms}^{\prime\prime} s^6 +\zeta_\mathrm{ms}^{\prime\prime\prime} s^8\right),\\
    \kappa_\mathrm{Oe}(s) =& \kappa_{\mathrm{Oe},0}\left(1+\zeta_\mathrm{Oe} s^2 +\zeta_\mathrm{Oe}^\prime s^4 + \zeta_\mathrm{Oe}^{\prime\prime} s^6 +\zeta_\mathrm{Oe}^{\prime\prime\prime} s^8\right).
\end{align}
In accordance with the experimental conditions, the magnetization of the polarizing layer is in-plane along $e_x$ and is partially tilted out-of-plane along $e_z$ by an applied perpendicular field $\mu_0H_{\mathrm{DC}}$
, following the approach of de Loubens \textit{et al.}~\cite{deLoubens2009}
Thus, both in-plane and out-of-plane components of the Slonczewski torques as well as the field-like torque are taken into account~\cite{Dussaux2012} and given by
\begin{align}
    \mathbf{F}_\mathrm{ST} =& \mathbf{F}_\mathrm{ST,slon}^\perp + \mathbf{F}_\mathrm{ST,slon}^\parallel + \mathbf{F}_\mathrm{ST, FLT},
\end{align}
where
\begin{align}
    \mathbf{F}_\mathrm{ST,slon}^\perp =& k_\mathrm{ST,\perp} J \left(\mathbf{e}_z\times\mathbf{X}\right),\\
    \mathbf{F}_\mathrm{ST,slon}^\parallel =& k_\mathrm{ST,\parallel} J\mathbf{e}_x,\\
    \mathbf{F}_\mathrm{ST, FLT} =& k_\mathrm{ST, FLT}J\mathbf{e}_y.
\end{align}

To reproduce experimental results, the RF bias excitation is introduced as an in-plane RF magnetic field along $\mathbf{e_x}$. 
In this model, the RF bias power $\mathrm{P}_\mathrm{bias}$ is set by an RF source with output impedance $Z_0$. Due to the impedance mismatch between the source and the field line of resistance $R_\mathrm{FL}$, only a portion of $\mathrm{P}_\mathrm{bias}$ is transmitted to the field line, corresponding to a dissipated power 
\begin{align}
    \mathrm{P}_\mathrm{FL} =& \frac{4 R_\mathrm{FL}Z_0}{(R_\mathrm{FL}+ Z_0)^2}\mathrm{P}_\mathrm{bias},
\end{align}
which in turn produces an in-plane RF magnetic field $\mu_0\mathbf{H}_x=\mu_0H_\mathrm{FL} \cos(\pi f_{\mathrm{IL}}^{\mathrm{sim}} t + \alpha) \mathbf{e_x}$ corresponding to the RF bias excitation, of RF magnetic field amplitude
\begin{align}
    \mu_0H_\mathrm{FL}  =& \sigma \sqrt{\mathrm{P}_\mathrm{FL}/R_\mathrm{FL}},
\end{align}
where $\sigma$ is the linear current-to-field conversion factor of the field line, set by its Ampère-Oersted field geometry.

We simulate the time evolution of the stochastic differential equation~\eqref{eq:TEA} using a second-order Heun scheme with a constant integration time step of $\Delta t= 1$~ps. 
Using down-conversion techniques employed in Ref.~\cite{phan2024unbiased}, we reconstruct the temporal evolution of the phase difference $\Delta \phi(t)$ from the $x(t)$ component, and identify, following the techniques described in Appendix~\ref{A.C}, the 0 and $\pi$ phase states and their corresponding mean dwell times. 
The system was simulated for durations of 10~ms (disregarding a 2~µs transient regime), obtained by concatenating five successive 2~ms simulations to record at least 650 switching events between the 0 and $\pi$ states, ensuring a statistically meaningful extraction of the corresponding mean dwell times.

The fitting procedure to identify the numerical values of all the above parameters is overparametrized, and there are multiple sets of parameters
that can describe the measured oscillator output. In this context, the parameters were adjusted heuristically so that the simulated free-running frequency and injection-locking range match their experimental counterparts. The numerical values of the constant parameters used to produce simulation results depicted in Fig.~\ref{fig:4}(b) and Fig.~\ref{fig:5}(b) are summarized in Table~\ref{tab:val_simu_param}.
\begin{table}
    \centering
	\begin{tabular*}{\columnwidth}{@{\extracolsep{\fill}}|lrl|}
		\hline 
		$G$ & -1.2772e-13 &\si{\kilo\gram\per\radian\per\second}\\
		$D$ & 3.0521e-15&\si{\kilo\gram\per\radian\per\second}\\
    	$k_{\mathrm{ms},0}$ & 2.2336e-04&\si{\kilo\gram\per\second\squared}\\
        $\zeta_\mathrm{ms}$ & 0.2055 &\\
    	$\zeta_\mathrm{ms}^\prime$ & 0.0791 &\\
    	$\zeta_\mathrm{ms}^{\prime\prime}$ & 0.0282 &\\
    	$\zeta_\mathrm{ms}^{\prime\prime\prime}$ & 0 &\\
    	$\kappa_{\mathrm{Oe},0}$ & -1.0842e-15 &\si{\joule\per\ampere}\\
    	$\zeta_\mathrm{Oe}$ & $-4/7$ & \\
    	$\zeta_\mathrm{Oe}^{\prime}$ & $-1/7$ & \\
    	$\zeta_\mathrm{Oe}^{\prime\prime}$ &  $-16/231$ & \\
    	$\zeta_\mathrm{Oe}^{\prime\prime\prime}$ & $-125/3003$ & \\
    	$k_{\mathrm{ST},\perp}$ & 1.2084e-16 &\si{\joule\per\ampere}\\
    	$k_{\mathrm{ST},\parallel}$ & -3.1222e-24 & \si{\joule\per\ampere}\\
    	$k_\mathrm{ST, FLT}$ & 1.8559e-23 & \si{\joule\per\ampere}\\
    	$\mu$ &-2.3713e-9 & \si{\ampere\meter}\\
        $T$ & 10 & \si{\kelvin}\\
        $R_0$ & 185 & nm\\
        $C$ & -1 & \\
        $\mu_0H_{\mathrm{DC, x}}$ & 3 & \si{\milli\tesla}\\
        $\mu_0H_{\mathrm{DC, y}}$ & 0 & \si{\milli\tesla}\\
        $\mu_0H_{\mathrm{DC, z}}$ & 355 & \si{\milli\tesla}\\
        $J_{\mathrm{DC}}$ & 4.09e10 & \si{\ampere\per\meter\squared}\\
        $J_{\mathrm{IL}}$ & 0.53e10 & \si{\ampere\per\meter\squared}\\
        $f_{\mathrm{IL}}^{\mathrm{sim}}$ & 516.7 & \si{\mega\hertz}\\
        $Z_0$ & 50 & \si{\ohm}\\
        $R_\mathrm{FL}$ & 6 & \si{\ohm}\\
        $\sigma$ & 0.05 & \si{\milli\tesla\per\milli\ampere}\\
		\hline
	\end{tabular*}
    \caption{Parameters used in the Thiele equation simulations}
    \label{tab:val_simu_param}
\end{table}

The simulation temperature is set to $T = 10$~K, significantly lower than the experimental temperature of 300~K. This reduced effective temperature is a known consequence of the Thiele equation being a collective-coordinate description with only two degrees of freedom. The fluctuation-dissipation relation applied to this reduced model does not quantitatively capture the noise amplitude of the full micromagnetic system at room temperature, and $T$ is therefore treated as an effective parameter adjusted to reproduce the experimentally observed average phase-jump rates~\cite{khalsa2015critical, grimaldi2014response}. Additionaly, the numerical value of $\sigma$ differs from the experimental value reported in Appendix~\ref{A.A} due to simplifications in the modeled field line geometry. 

\section{Analytical calculation for the toy model} \label{A.E}
In this appendix, we detail and generalize the model of Sec.~\ref{V} to $n$\textsuperscript{th}-harmonic injection locking ($n$HIL), where $n = p/q$ with $p, q$ positive integers ($n \in \mathbb{Q}^{+*}$), covering both integer and fractional synchronization~\cite{LebrunPRL, Urazhdin2010Fractional}. A weak RF bias excitation is applied at $f_\mathrm{bias} = f_\mathrm{IL}/n$.
We assume that temporal amplitude variations around the stationary oscillation amplitude are negligible and can therefore be absorbed into effective model parameters. Under this assumption, amplitude-phase coupling in the SHIL regime mainly renormalizes the coupling coefficient $\Omega$ and the phase offset, without changing the qualitative form of the phase equation.
In this case, the phase dynamics reduce to a modified Adler equation~\cite{adler1946, slavin_nonlinear_2009} for the phase difference $\Delta\phi = \phi_\mathrm{STNO} - \phi_\mathrm{IL}/n$,
\begin{equation}
    \dfrac{d\Delta\phi}{dt} = -\dfrac{\delta}{n} - \dfrac{\Omega}{n}\sin\left(n\Delta\phi\right) - \Omega'\sin\left(\Delta\phi - \alpha + \alpha_0\right).
    \label{eq:A.E.1}
\end{equation}
The first term contains the frequency detuning $\delta = 2\pi(nf_\mathrm{STNO} - f_\mathrm{IL})$ between the oscillator and the $n$HIL drive. The second term describes the $n$HIL synchronization, with $\Omega$ the associated locking range, defined as the maximum frequency detuning $|\delta|$ before losing synchronization and referenced to the source frequency $f_\mathrm{IL}$. The third term is the perturbation introduced by the RF bias excitation, with $\Omega'$ its locking range, defined analogously and referenced to the source frequency $f_\mathrm{bias}$, $\alpha = \phi_\mathrm{bias} - \phi_\mathrm{IL}/n$ the relative phase of the bias with respect to the $n$HIL drive, and $\alpha_0$ an effective phase offset arising from the oscillator nonlinearity and experimental conditions.
Integrating Eq.~\eqref{eq:A.E.1} yields the phase quasipotential $V(\Delta\phi)$, defined by $d\Delta\phi/dt = -dV/d\Delta\phi$,
\begin{equation}
    V(\Delta\phi) = \dfrac{\delta}{n}\Delta\phi - \dfrac{\Omega}{n^2}\cos(n\Delta\phi) - \Omega'\cos(\Delta\phi - \alpha + \alpha_0).
    \label{eq:A.E.2}
\end{equation}
For $n=2$, Eq.~\eqref{eq:A.E.2} recovers the quasipotential of Eq.~\eqref{eq:potential} used in the main text. Note that Eq.~\eqref{eq:A.E.1} and Eq.~\eqref{eq:A.E.2} are conventionally written in terms of $\psi = n\phi_\mathrm{STNO} - \phi_\mathrm{IL}$, which takes the injection-locking drive as the phase reference~\cite{hem2019power}. We instead use $\Delta\phi = \phi_\mathrm{STNO} - \phi_\mathrm{IL}/n$ throughout, as it is the phase difference directly accessible in the experiment.



Eq.~\eqref{eq:A.E.2} does not admit a closed-form solution in general. We therefore apply first-order classical perturbation theory~\cite{nayfeh2011introduction, strogatz2018nonlinear} to derive approximate expressions for the equilibrium points $\Delta\phi^{eq}$ and the barrier heights $\Delta V$ separating them. The quasipotential is decomposed as
\begin{equation}
    \begin{aligned}
        V(\Delta\phi) &= V_\mathrm{IL}(\Delta\phi) + \epsilon\, V_\mathrm{bias}(\Delta\phi) + \mathcal{O}(\epsilon^2) \\
        &= \dfrac{\delta}{n}\Delta\phi - \dfrac{\Omega}{n^2}\cos(n\Delta\phi) \\
        &\quad - \epsilon\,\Omega\cos(\Delta\phi - \alpha + \alpha_0) + \mathcal{O}(\epsilon^2),
    \end{aligned}
    \label{eq:A.E.3}
\end{equation}
where $V_\mathrm{IL}$ corresponds to the phase quasipotential for zero bias excitation, which admits a known analytical solution~\cite{slavin_nonlinear_2009, phan2024unbiased}, and $V_\mathrm{bias}$ is the perturbation introduced by the RF bias excitation. The perturbation parameter $\epsilon = \Omega'/\Omega$ is assumed small ($\epsilon \ll 1$), and the expansion is carried out to first order in $\epsilon$.
The equilibrium points $V'(\Delta\phi)=0$ can then be approximated as 
\begin{equation}
    \Delta\phi^{eq} = \Delta\phi^{eq}_0 + \epsilon\Delta\phi^{eq}_1 + \mathcal{O}\left(\epsilon^2\right).
    \label{eq:A.E.4}
\end{equation}
Here $\Delta\phi^{eq}_0$ is the unperturbed equilibrium point ($\epsilon = 0$), and $\Delta\phi^{eq}_1$ is the first-order correction. Substituting Eq.~\eqref{eq:A.E.4} into $V'(\Delta\phi) = 0$ and expanding to first order in $\epsilon$, we obtain, with $\theta = \arcsin(\delta/\Omega)$ as defined in Sec.~\ref{V} and $k$ an integer ($k \in \mathbb{Z}$) indexing the periodic equilibria,
\begin{align}
    & \Delta\phi^{eq}_{0,\mathrm{stable}} = \dfrac{-\theta + 2k\pi}{n}, \label{eq:A.E.5} \\
    & \Delta\phi^{eq}_{0,\mathrm{unstable}} = \dfrac{\pi + \theta + 2k\pi}{n}, \label{eq:A.E.6} \\
    & \Delta\phi^{eq}_1 = -\dfrac{\sin(\Delta\phi^{eq}_0 - \alpha + \alpha_0)}{\cos(n\Delta\phi^{eq}_0)}. \label{eq:A.E.7}
\end{align}
The zeroth-order terms correspond to the unperturbed stable ($V_\mathrm{IL}''(\Delta\phi^{eq}_{0,\mathrm{stable}}) > 0$) and unstable ($V_\mathrm{IL}''(\Delta\phi^{eq}_{0,\mathrm{unstable}}) < 0$) equilibria, while the first-order correction $\Delta\phi^{eq}_1$ depends only on the phase offsets $\alpha$ and $\alpha_0$ and is independent of $\Omega'$. Inserting Eq.~\eqref{eq:A.E.4} into Eq.~\eqref{eq:A.E.3} and expanding to first order in $\epsilon$ yields the quasipotential at the equilibria,
\begin{equation}
    \begin{aligned}
      V(\Delta\phi^{eq}) \approx \dfrac{\delta}{n}\Delta\phi^{eq}_0 - \dfrac{\Omega}{n^2}\cos(n\Delta\phi^{eq}_0) \\
     -\Omega'\cos(\Delta\phi^{eq}_0 - \alpha + \alpha_0).
    \end{aligned} \label{eq:A.E.8}
\end{equation}
Notably, to first order in $\epsilon$, the quasipotential at the equilibria is independent of the correction $\Delta\phi^{eq}_1$. Substituting the stable and unstable zeroth-order equilibria of Eqs.~\eqref{eq:A.E.5} and~\eqref{eq:A.E.6} into Eq.~\eqref{eq:A.E.8} yields
\begin{align}
    \begin{split}
        V(\Delta\phi^{eq}_\mathrm{stable}) \approx & \dfrac{\delta}{n}\left(\dfrac{-\theta + 2k\pi}{n}\right) - \dfrac{\Omega}{n^2}\cos\theta \\
        & - \Omega'\cos\!\left(\dfrac{-\theta + 2k\pi}{n} - \alpha + \alpha_0\right),
    \end{split} \label{eq:A.E.9} \\
    \begin{split}
        V(\Delta\phi^{eq}_\mathrm{unstable}) \approx & \dfrac{\delta}{n}\left(\dfrac{\pi + \theta + 2k\pi}{n}\right) + \dfrac{\Omega}{n^2}\cos\theta \\
        & - \Omega'\cos\!\left(\dfrac{\pi + \theta + 2k\pi}{n} - \alpha + \alpha_0\right).
    \end{split} \label{eq:A.E.10}
\end{align}
The quasipotential barriers separate adjacent stable and unstable equilibrium points, with $\Delta V = V(\Delta\phi^{eq}_\mathrm{unstable}) - V(\Delta\phi^{eq}_\mathrm{stable})$. We distinguish two families of barriers according to whether overcoming them results in an increase or a decrease of $\Delta\phi$, denoted as $\Delta V_\mathrm{in}(k)$ and $\Delta V_\mathrm{de}(k)$ respectively,
\begin{align}
    \Delta V_\mathrm{in}(k) &\approx \Delta V_\mathrm{in}^0 + 2\Omega' \sin\!\left(\tfrac{2\theta + \pi}{2n}\right) \sin\!\left(\tfrac{(4k+1)\pi}{2n} - \alpha + \alpha_0\right), \label{eq:A.E.11} \\
    \Delta V_\mathrm{de}(k) &\approx \Delta V_\mathrm{de}^0 + 2\Omega' \sin\!\left(\tfrac{2\theta - \pi}{2n}\right) \sin\!\left(\tfrac{(4k-1)\pi}{2n} - \alpha + \alpha_0\right), \label{eq:A.E.12}
\end{align}
where $\Delta V_\mathrm{in}^0=( (2\theta+\pi)\delta + 2\Omega\cos\theta)/n^2$, and $\Delta V_\mathrm{de}^0=( (2\theta-\pi)\delta + 2\Omega\cos\theta)/n^2$ are the corresponding barrier heights in the absence of RF bias excitation. 
For $n=2$, Eq.~\eqref{eq:A.E.11} and Eq.~\eqref{eq:A.E.12} reduce to the four barriers of Eq.~\eqref{eq:barriers}, with $\Delta V_{\mathrm{in}}(k=0) = \Delta V_1$, $\Delta V_{\mathrm{in}}(k=1) = \Delta V_3$, $\Delta V_{\mathrm{de}}(k=1) = \Delta V_2$, and $\Delta V_{\mathrm{de}}(k=0) = \Delta V_4$.

\section{Markov model analysis for the toy model} \label{A.F}

In this appendix, we derive the mean dwell times and stationary occupation probabilities of the 0 and $\pi$ phase states from the four-state Markov model introduced in Sec.~\ref{V} (see Fig.~\ref{fig:6}). The four states A, B, C, D correspond to successive minima of the quasipotential, with A and C identified with $\Delta\phi^{eq}_\mathrm{stable}(2k)$ and B and D with $\Delta\phi^{eq}_\mathrm{stable}(2k+1)$ for integer $k$. The 0 and $\pi$ phase states observed experimentally correspond to the macrostates $\{\mathrm{A}, \mathrm{C}\}$ and $\{\mathrm{B}, \mathrm{D}\}$, respectively. Restricting the dynamics to nearest-neighbor jumps, the $2\pi$-periodicity of the quasipotential reduces the eight possible jumps to four distinct phase jump rates $r_1, r_2, r_3, r_4$ (see Fig.~\ref{fig:6}~(b)). Treating each jump as a Poisson process activated by coupling to the thermal bath, the master equation $\dot{\bm{P}} = \mathbf{R}\bm{P}$ for the microstate probabilities $\bm{P} = (P_\mathrm{A}, P_\mathrm{B}, P_\mathrm{C}, P_\mathrm{D})^\top$ involves the rate matrix
\begin{equation}
    \mathbf{R} = \begin{pmatrix}
        -(r_1 + r_4) & r_2 & 0 & r_3 \\
        r_1 & -(r_2 + r_3) & r_4 & 0 \\
        0 & r_3 & -(r_1 + r_4) & r_2 \\
        r_4 & 0 & r_1 & -(r_2 + r_3)
    \end{pmatrix}.
    \label{eq:A.F.1}
\end{equation}
Solving $\dot{\bm{P}} = \mathbf{0}$ yields the stationary microstate probabilities, with $P_\mathrm{A} = P_\mathrm{C}$ and $P_\mathrm{B} = P_\mathrm{D}$ enforced by the $2\pi$-periodicity. Summing within each macrostate, $P_0 = P_\mathrm{A} + P_\mathrm{C}$ and $P_\pi = P_\mathrm{B} + P_\mathrm{D}$, leads to the closed form
\begin{equation}
    P_0 = \dfrac{r_2 + r_3}{r_1 + r_2 + r_3 + r_4}, \qquad
    P_\pi = \dfrac{r_1 + r_4}{r_1 + r_2 + r_3 + r_4}.
    \label{eq:A.F.2}
\end{equation}

Equivalently, the master equation reduces to a two-state problem for the macrostate probabilities, with total escape rates $r_1 + r_4$ out of the 0 state (via the A$\to$B and C$\to$D transitions) and $r_2 + r_3$ out of the $\pi$ state (via the B$\to$A and D$\to$C transitions). The mean dwell times are then the inverse of these total escape rates,
\begin{equation}
    \tau_0 = \dfrac{1}{r_1 + r_4}, \qquad
    \tau_\pi = \dfrac{1}{r_2 + r_3},
    \label{eq:A.F.3}
\end{equation}
recovering Eq.~\eqref{eq:dwell_times} of the main text. Combining Eqs.~\eqref{eq:A.F.2} and~\eqref{eq:A.F.3} gives the expected relation $P_{0,\pi} = \tau_{0,\pi}/(\tau_0 + \tau_\pi)$ used in Sec.~\ref{III}.
Combined with the Arrhenius rates of Eq.~\eqref{eq:rate}, Eq.~\eqref{eq:A.F.3} provides the explicit dependence of $\tau_0$ and $\tau_\pi$ on the quasipotential barriers $\Delta V_i$.


\section{Details on the validation of the toy model against mean dwell times} \label{A.G}


To assess the predictive power of the analytical model developed in Appendices~\ref{A.E} and~\ref{A.F}, we relate the experimentally accessible mean dwell times of the biased SHIL STNO to the corresponding quasi-potential barriers derived from $V(\Delta\phi)$. We then jointly fit the experimentally measured and simulated dwell times $\tau_0$ and $\tau_\pi$ using Eqs.~\eqref{eq:dwell_times} and~\eqref{eq:A.F.3}, combined with the Arrhenius-type rates of Eq.~\eqref{eq:rate} and the barrier heights of Eq.~\eqref{eq:barriers}.
Each fit is performed simultaneously on $\tau_0$ and $\tau_\pi$ and across all values of the swept parameter (either the bias phase $\alpha$ or the bias power $\mathrm{P}_\mathrm{bias}$). Representative fits are shown as solid lines in Fig.~\ref{fig:4}, with the corresponding fitting parameters reported in Tables~\ref{tab:A.G.2} and~\ref{tab:A.G.3}.
Table~\ref{tab:A.G.1} lists the independently measured or estimated quantities used to set the initial guesses and boundaries of the fitting parameters. Asterisks denote order-of-magnitude estimates.

\begin{table} \centering
    \begin{tabular}{cc|c|c} 
         \textbf{Symbol} & \textbf{Unit} & \textbf{Experiments} & \textbf{Simulations}\\ \hline
        
        $\Omega^\prime$(-40 dBm) & MHz & 2.76 & 10.68 \\
        
        $f_p$ & MHz & 13.15 & 1.12 \\
        
        $\eta$ & Hz/J & 2.57$\times10^{26}$* & 1.41$\times10^{28}$*  \\ \hline
    \end{tabular}
    \caption{Parameters used as estimation of initial guesses or boundaries in the fit of experimental and simulated stable phase states dwell times results using Eq.~\eqref{eq:dwell_times}, respectively for each column. Used for the fits shown in Fig.~\ref{fig:4} and~\ref{fig:5}.}
    \label{tab:A.G.1}
\end{table}
\begin{table*}[ht] \centering
    \begin{tabular}{ccc|ccc|ccc} 
        \multicolumn{9}{c}{\textbf{Toy model parameters used to fit experiments using Eq.~ \eqref{eq:dwell_times}}} \\ \hline
        
        \multicolumn{9}{c}{\textbf{Constant parameters} \qquad\qquad
            \begin{tabular}{c @{ = } c}
                $n$ & 2 \\ $\delta$ & -3.02 MHz \\ $\Omega$ & 21.86 MHz \\ $T$ & 300 K
            \end{tabular}} \\ \hline\hline
        
        \multicolumn{9}{c}{\textbf{Fitting parameters}} \\ \hline
        
        \multicolumn{3}{c|}{\textbf{$\tau$ vs. $\alpha$ for $P_\mathrm{bias}=-50$ dBm}} & 
        \multicolumn{3}{c|}{\textbf{$\tau$ vs. $P_\mathrm{bias}$ for $\alpha=\pi/9$ rad}} & 
        \multicolumn{3}{c}{\textbf{$\tau$ vs. $P_\mathrm{bias}$ for $\alpha=\pi$ rad}} \\
        
        \textbf{Symbol} & \textbf{Value} & \textbf{Unit} & \textbf{Symbol} & \textbf{Value} & \textbf{Unit} & \textbf{Symbol} & \textbf{Value} & \textbf{Unit} \\ \hline
        
        $\Omega^\prime$ & 1.12 $\pm$ 0.12 & MHz & 
        $\alpha$ & 0.52 & rad & 
        $\alpha$ & 3.49 & rad \\
        
        $\alpha_0$ & 0.24 $\pm$ 0.05 & rad & 
        $\alpha_0$ & 0.11 & rad & 
        $\alpha_0$ & 0.12 & rad \\
        
        $\eta$ & 6.13$\times10^{26}$ $\pm$ 5.13$\times10^{25}$ & Hz/J & 
        $\eta$ & 6.25$\times10^{26}$ & Hz/J & 
        $\eta$ & 6.25$\times10^{26}$ & Hz/J \\
        
        $r_0(0)$ & 11.7 $\pm 3.64$ & MHz & 
        $r_0(0)$ & 9.76 & MHz & 
        $r_0(0)$ & 10.76 & MHz \\
        
        $r_0(\pi)$ & 12.05 $\pm 3.74$ & MHz & 
        $r_0(\pi)$ & 11.11 & MHz & 
        $r_0(\pi)$ & 11.31 & MHz
        \\ \hline
    \end{tabular}
    \caption{Parameters used to fit experimental stable phase states dwell times results using Eq.~\eqref{eq:dwell_times}, respectively for each column in Fig.~\ref{fig:4}~(a),~(b)~and~(c). For the mean dwell times fitting versus bias power $P_\mathrm{bias}$, no fitting errors could be estimated. It is mostly due to the model being over-parametrized and the small number of fitted points compared to the number of fitting parameters.}
    \label{tab:A.G.2}
\end{table*}

\begin{table*} \centering
    \begin{tabular}{ccc|ccc|ccc} 
        \multicolumn{9}{c}{\textbf{Toy model parameters used to fit numerical simulations using Eq.~\eqref{eq:dwell_times}}} \\ \hline
        
        \multicolumn{9}{c}{\textbf{Constant parameters} \qquad\qquad
            \begin{tabular}{c @{ = } c}
                $n$ & 2 \\ $\delta$ & -16.96 MHz \\ $\Omega$ & 30.16 MHz \\ $T$ & 10 K
            \end{tabular}} \\ \hline\hline
        
        \multicolumn{9}{c}{\textbf{Fitting parameters}} \\ \hline
        
        \multicolumn{3}{c|}{\textbf{$\tau$ vs. $\alpha$ for $P_\mathrm{bias}=-50$ dBm}} & 
        \multicolumn{3}{c|}{\textbf{$\tau$ vs. $P_\mathrm{bias}$ for $\alpha=\pi/9$ rad}} & 
        \multicolumn{3}{c}{\textbf{$\tau$ vs. $P_\mathrm{bias}$ for $\alpha=\pi$ rad}} \\
        
        \textbf{Symbol} & \textbf{Value} & \textbf{Unit} & \textbf{Symbol} & \textbf{Value} & \textbf{Unit} & \textbf{Symbol} & \textbf{Value} & \textbf{Unit} \\ \hline
        
        $\Omega^\prime$ & 3.32 $\pm$ 0.42 & MHz & 
        $\alpha$ & 0 $\pm $182499.88 & rad & 
        $\alpha$ & 3.16 $\pm$216.33 & rad \\
        
        $\alpha_0$ & 1.68 $\pm$0.08 & rad & 
        $\alpha_0$ & 3.02 $\pm$74310.36 & rad & 
        $\alpha_0$ & 2.39 $\pm$67.71 & rad \\
        
        $\eta$ & 2.34$\times10^{28}$ $\pm$1.54$\times10^{27}$ & Hz/J & 
        $\eta$ & 2.46$\times10^{28}$ $\pm$3.54$\times10^{27}$ & Hz/J & 
        $\eta$ & 2.46$\times10^{28}$ $\pm$3.54$\times10^{27}$ & Hz/J \\
        
        $r_0(0)$ & 1.1 $\pm$0.14 & MHz & 
        $r_0(0)$ & 1.03 $\pm$0.41 & MHz & 
        $r_0(0)$ & 0.78 $\pm$0.23 & MHz \\
        
        $r_0(\pi)$ & 1.1 $\pm$0.14 & MHz & 
        $r_0(\pi)$ & 1.03 $\pm$0.43 & MHz & 
        $r_0(\pi)$ & 1.06 $\pm$0.39 & MHz
        \\ \hline
    \end{tabular}
    \caption{Parameters used to fit numerically simulated stable phase states dwell times results using Eq.~\eqref{eq:dwell_times}, respectively for each column in Fig.~\ref{fig:4}~(d),~(e)~and~(f).}
    \label{tab:A.G.3}
\end{table*}

The locking range $\Omega'$ due to the RF bias excitation at $\mathrm{P}_\mathrm{bias} = -40$~dBm was extracted from the Lorentzian fitting procedure described in Appendix~\ref{A.B}, both for experiments and simulations. At lower powers, the locking range becomes too narrow to be reliably extracted from the spectra. For these powers, we therefore use the scaling relation $\Omega'(\mathrm{P}_\mathrm{bias}) \approx \Omega'(-40~\mathrm{dBm}) \times 10^{(\mathrm{P}_\mathrm{bias} + 40)/20}$, which follows from $\Omega'$ being proportional to the RF magnetic field amplitude~\cite{slavin_nonlinear_2009}, itself proportional to the square root of the source power.

The power relaxation frequency $f_p$~\cite{slavin_nonlinear_2009} is estimated from the Fourier-transform analysis of the amplitude and phase noise~\cite{quinsat2010phasenoise, grimaldi2014response}. In our interpretation, this frequency sets an upper bound for the effective attempt rate $r_0$ in Eq.~\eqref{eq:rate}, since the phase cannot attempt jumps faster than the intrinsic response rate of the oscillator.
The scaling factor $\eta$ quantifies the coupling between the thermal energy $k_BT$ and the oscillator's phase fluctuations. An order-of-magnitude estimate is obtained in the free-running regime from the relation $\Delta f = \eta k_BT$ between the thermal contribution to the frequency linewidth $\Delta f$ and the thermal energy~\cite{slavin_nonlinear_2009, grimaldi2014response}, which can also be interpreted as a damping-like torque efficiency. Since the frequency linewidth captures only part of the full noise spectrum of the oscillator, the values in Table~\ref{tab:A.G.1} are lower-bound estimates and are used only to set the order of magnitude of $\eta$ in the fits. A single value of $\eta$ is shared across all jointly fitted curves.

Tables~\ref{tab:A.G.2} and~\ref{tab:A.G.3} report the fitted parameters and their uncertainties for the experimental and simulated dwell times, respectively, corresponding to the fits shown in Fig.~\ref{fig:4}. For the sweeps of $\mathrm{P}_\mathrm{bias}$ in the experimental fits (Table~\ref{tab:A.G.2}), the covariance matrix could not be reliably estimated due to the limited number of data points relative to the number of free parameters, and uncertainties are therefore omitted in this case.

%

\end{document}